\documentclass[iop,revtex4]{emulateapj} \newcommand{\ltab}{\LongTables}

\usepackage{hyperref}

\hypersetup{
    bookmarks=true,         
    unicode=false,          
    pdftoolbar=true,        
    pdfmenubar=true,        
    pdffitwindow=true,     
    pdfstartview={FitH},    
    pdftitle={HST CEMP-s},    
    pdfauthor={vmplacco},     
    pdfsubject={Astronomy},   
    pdfcreator={dvipdf},   
    pdfproducer={dvipdf}, 
    pdfkeywords={metal-poor stars},
    pdfnewwindow=true,      
    colorlinks=true,       
    linkcolor=red,          
    citecolor=blue,        
    filecolor=magenta,      
    urlcolor=cyan,           
    breaklinks=true,
    linktocpage
}

\newcommand{\bd}{\object{BD+44$^\circ$493}}
\newcommand{\hda}{\object{HD~196944}}
\newcommand{\hdb}{\object{HD~201626}}
\newcommand{\eps}[1]{\mbox{log~$\epsilon$\,(#1)}}
\newcommand{\metal}{[Fe/{}H]}
\newcommand{\cfe}{[C/{}Fe]}
\newcommand{\abund}[2]{[#1/{}#2]}

\newcommand{\teff}{$T_{\rm eff}$\,}
\newcommand{\logg}{log\,$g$\,}
\newcommand{\xx}{{\tablenotemark{a}}}
\newcommand{\yy}{{\tablenotemark{b}}}
\newcommand{\zz}{{\tablenotemark{c}}}
\input{colordvi}

\shorttitle{HST Spectra of CEMP-$s$ Stars}
\shortauthors{Placco et al.}

\begin{document}

\title{\textit{Hubble Space Telescope} Near-Ultraviolet \\
Spectroscopy of Bright CEMP-$s$ Stars\footnotemark[1]}

\footnotetext[1]{
The data presented herein were obtained with the (i) NASA/ESA Hubble Space
Telescope, obtained at the Space Telescope Science Institute, which is operated
by the Association of Universities for Research in Astronomy, Inc., under NASA
contract NAS~5-26555. (These observations are associated with program GO-12554,
datasets OBQ601010-30 and OBQ602010-30.); and (ii) W.~M.\ Keck Observatory,
which is operated as a scientific partnership among the California Institute of
Technology, the University of California and the National Aeronautics and Space
Administration. (The Observatory was made possible by the generous financial
support of the W.~M.\ Keck Foundation).}

\author{
Vinicius M. Placco\altaffilmark{2,3},
Timothy   C. Beers\altaffilmark{2,3},
Inese     I. Ivans\altaffilmark{4},
Dan         Filler\altaffilmark{4},
Julie A.      Imig\altaffilmark{4},\\
Ian    U. Roederer\altaffilmark{5,3},
Carlo        Abate\altaffilmark{6},
Terese      Hansen\altaffilmark{7},
John      J. Cowan\altaffilmark{8},
Anna        Frebel\altaffilmark{9},\\
James    E. Lawler\altaffilmark{10},
Hendrik     Schatz\altaffilmark{11,3},
Christopher Sneden\altaffilmark{12},
Jennifer S. Sobeck\altaffilmark{13},\\
Wako          Aoki\altaffilmark{14},
Verne     V. Smith\altaffilmark{15},
Michael      Bolte\altaffilmark{16}
}

\altaffiltext{2}{Department of Physics, University of Notre Dame, 
                 Notre Dame, IN 46556, USA}
\altaffiltext{3}{JINA Center for the Evolution of the Elements, USA}
\altaffiltext{4}{Department of Physics and Astronomy, The University of Utah,
                 Salt Lake City, UT 84112, USA}
\altaffiltext{5}{Department of Astronomy, University of Michigan,
                 Ann Arbor, MI 48109, USA}
\altaffiltext{6}{Argelander Institut f\"ur Astronomie, Auf dem H\"ugel 71, 
                 53121, Bonn, Germany}
\altaffiltext{7}{Landessternwarte, ZAH, K\"onigstuhl 12, 69117 Heidelberg, 
                 Germany}
\altaffiltext{8}{Homer L. Dodge Department of Physics and Astronomy, University
                 of Oklahoma, Norman, OK 73019, USA}
\altaffiltext{9}{Kavli Institute for Astrophysics and Space Research and
                 Department of Physics, Massachusetts Institute of Technology, 
				 Cambridge, MA 02139, USA}
\altaffiltext{10}{Department of Physics, University of Wisconsin, 
                  Madison, WI 53706, USA}                                                
\altaffiltext{11}{National Superconducting Cyclotron Laboratory, Michigan State 
                  University, East Lansing, MI 48824, USA} 
\altaffiltext{12}{Department of Astronomy and McDonald Observatory, University
                  of Texas, Austin, TX 78712, USA}
\altaffiltext{13}{Department of Astronomy, University of Virginia,
                  Charlottesville, VA 22904, USA}
\altaffiltext{14}{National Astronomical Observatory of Japan, 2-21-1 Osawa,
                  Mitaka, Tokyo 181-8588, Japan}
\altaffiltext{15}{National Optical Astronomy Observatory, Tucson, AZ 85719, USA}
\altaffiltext{16}{Department of Astronomy and Astrophysics, University of 
                  California, Santa Cruz, CA 95064, USA}

\addtocounter{footnote}{16}

\begin{abstract}

We present an elemental-abundance analysis, in the near-ultraviolet (NUV)
spectral range, for the bright carbon-enhanced metal-poor (CEMP) stars \hda{}
($V = 8.40$, [Fe/H] $= -2.41$)  and \hdb{} ($V = 8.16$, [Fe/H] = $-1.51$),
based on data acquired with the Space Telescope Imaging Spectrograph (STIS) on
the \textit{Hubble Space Telescope}. Both of these stars belong to the
sub-class CEMP-$s$, and exhibit clear over-abundances of heavy elements
associated with production by the slow neutron-capture process. \hda{} has been
well-studied in the optical region, but we add abundance results
for six species (Ge, Nb, Mo, Lu, Pt, and Au) that are only accessible in the
NUV. In addition, we provide the first determination of its orbital period,
P=1325~days. \hdb{} has only a limited number of abundance results based on
previous optical work -- here we add five new species from the NUV,
including Pb.  We compare these results with models of binary-system
evolution and $s$-process element production in stars on the asymptotic
giant branch, with the goal of explaining their origin and evolution. 
Our best-fitting models
for \hda{}  ($M_{1,i}=0.9 M_{\odot}$, $M_{2,i}=0.86 M_{\odot}$, for \metal=$-2.2$),
and  \hdb{} ($M_{1,i}=0.9 M_{\odot}$, $M_{2,i}=0.76 M_{\odot}$, for \metal=$-2.2$;
             $M_{1,i}=1.6 M_{\odot}$, $M_{2,i}=0.59 M_{\odot}$, for \metal=$-1.5$)
are consistent with the current accepted scenario for the formation of CEMP-$s$
stars.

\end{abstract}


\section{Introduction}
\label{intro}

Carbon-enhanced metal-poor (CEMP) stars have received increased attention in
the recent literature due to their clear importance as probes of a number of
astrophysical phenomena, e.g., the production of elements by the first
generations of stars in the universe \citep[][and references
therein]{beers2005,frebel2015}, the mass-transfer process in binary systems
\citep{abate2013,abate2015a,abate2015b,abate2015c}, and the nature of neutron-capture
processes responsible for the production of elements beyond the iron peak
\citep{bisterzo2010,bisterzo2011,bisterzo2012}. The CEMP class of stars
comprises a number of sub-classes \citep[originally defined by][but modified
somewhat by subsequent authors]{beers2005}, based on the abundances of their
neutron-capture elements: (1) CEMP-no stars, which exhibit no over-abundances
of neutron-capture elements, (2) CEMP-$s$ stars, which exhibit neutron-capture
over-abundances consistent with the slow neutron-capture process, (3) CEMP-$r$
stars, with neutron-capture over-abundances associated with the rapid neutron-capture
process, and (4) CEMP-$r/s$ stars, which exhibit neutron-capture over-abundances
that suggest contribution from both the slow and rapid neutron-capture
processes. 

The great majority of abundance studies for CEMP stars have been restricted to
the optical region, because samples of C-enhanced stars that are sufficiently
bright to be observed at high spectral resolution in the near-ultraviolet (NUV)
from space are extremely limited.  \citet{sneden2003}, \citet{cowan2005}, and
\citet{roederer2009} studied the STIS spectrum of CS~22892$-$052, a CEMP-$r$
star. For CEMP-no stars, the first and only such study was the
\citet{placco2014b} analysis of the bright ($V = 9.1$), extremely metal-poor
([Fe/H] = $-3.8$) CEMP-no star \bd. The authors showed that the abundances of
the elements beryllium (Be, $Z$=4) and boron (B, $Z$=5) -- thought to originate
from cosmic-ray spallation reactions -- are at the lowest level yet observed
among all very and extremely metal-poor stars to date (\eps{Be} $< -2.3$ and
\eps{B} $< -0.70$, respectively). Their derived upper limit on the abundance of
lead (Pb; $Z$=82; \eps{Pb} $< -0.23$) is difficult to reconcile with
$s$-process nucleosythesis in low-mass asymptotic giant-branch (AGB) stars with
the highest neutron exposures possible.  Both of these results strengthen the
argument that \bd, and by implication other CEMP-no stars, could well be
bona-fide second-generation stars, born from an interstellar medium polluted by
massive first-generation stars \citep[see also][for a recent
review]{frebel2015}.   

In this study we continue our examination of the elemental-abundance patterns
for CEMP stars, supplementing studies in the optical region with new NUV
\textit{Hubble Space Telescope} (\textit{HST}) Space Telescope Imaging
Spectrograph (STIS) spectroscopic data for the bright, very metal-poor
CEMP-$s$ stars \hda{} and \hdb. We obtain abundances or upper limits for a
number of elements that are challenging or impossible to obtain from
ground-based studies, but are nevertheless important for constraining
detailed predictions of their production by AGB stars.  These elements
include: carbon (C; $Z$=6), oxygen (O; $Z$=8), titanium (Ti; $Z$=22),
chromium (Cr; $Z$=24), manganese (Mn; $Z$=25), nickel (Ni; $Z$=28),
germanium (Ge; $Z$=32), zirconium (Zr; $Z$=40), niobium (Nb; $Z$=41),
molybdenum (Mo; $Z$=42), cadmium (Cd; $Z$=48), lutetium (Lu; $Z$=71),
hafnium (Hf; $Z$=72), osmium (Os; $Z$=76), platinum (Pt; $Z$=78), gold (Au;
$Z$=79), and lead (Pb; $Z=82$).  Section~\ref{obssec} describes our
observations, data reduction, atmospheric parameter determinations, and
radial-velocity variations.  Section~\ref{absec} describes our abundance
analysis in detail, followed by a comparison with theoretical AGB models in
Section~\ref{diss}. We present a brief discussion and our conclusions in
Section~\ref{conc}.
 
\section{Observations and Measurements} \label{obssec}

We studied high-resolution spectra of \hda{} and \hdb{} from the NUV to optical
wavelengths, employing data gathered with the HST/STIS and Keck/HIRES
spectrographs. Below we provide a description of the observations, data
reduction, and our model-atmosphere parameter determinations.

\subsection{\textit{HST}/STIS Spectra}

STIS \citep{kimble1998,woodgate1998} observations of \hda{} and \hdb{} were
obtained as part of Program GO-12554, using the E230M echelle grating, centered
at 2707\,{\AA}, and the NUV Multianode Microchannel Array (MAMA) detector.
There was one observational sequence of three individual exposures for each
star, taken on 2012 April 29 (\hdb) and 2012 September 22 (\hda). The total
integration time was about 7.9~ks per star.  The 0\farcs06\,$\times$\,0\farcs2
slit yields a resolving power of R~$\sim$~30,000. Our setup produced a
wavelength coverage from 2280\,{\AA}--3070\,{\AA} in a single exposure. The
observations were reduced and calibrated using the standard \textit{calstis}
pipeline. The S/N of the combined spectrum varies from $\sim$45~pix$^{-1}$ near
2300\,{\AA}, to $\sim$70~pix$^{-1}$ near 2700\, {\AA}, to $>$90~pix$^{-1}$ near
3070\, {\AA}.

\begin{figure}[!ht]
\epsscale{1.15}
\plotone{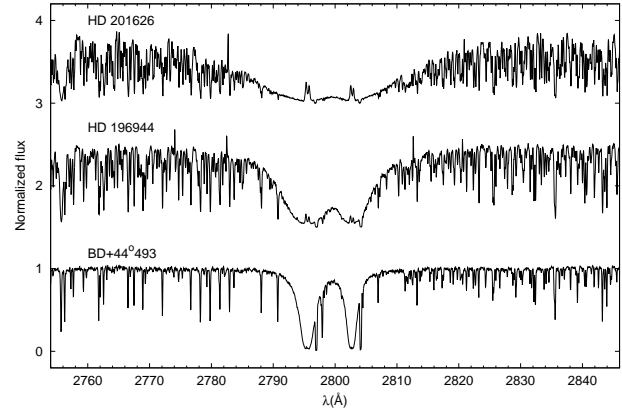}
\caption{\textit{HST}/STIS NUV spectra for \protect\hdb{} (\metal=$-1.51$) and
\protect\hda\ (\metal=$-2.41$), compared with the spectrum of the EMP
\protect\bd\ (\metal=$-3.80$). The region shown includes the \ion{Mg}{2} doublet at
2800\,\AA. These stars have similar \teff; the differences in
metallicity are apparent.}
\label{mg_comp}
\end{figure}

\begin{figure}[!ht]
\epsscale{1.15}
\plotone{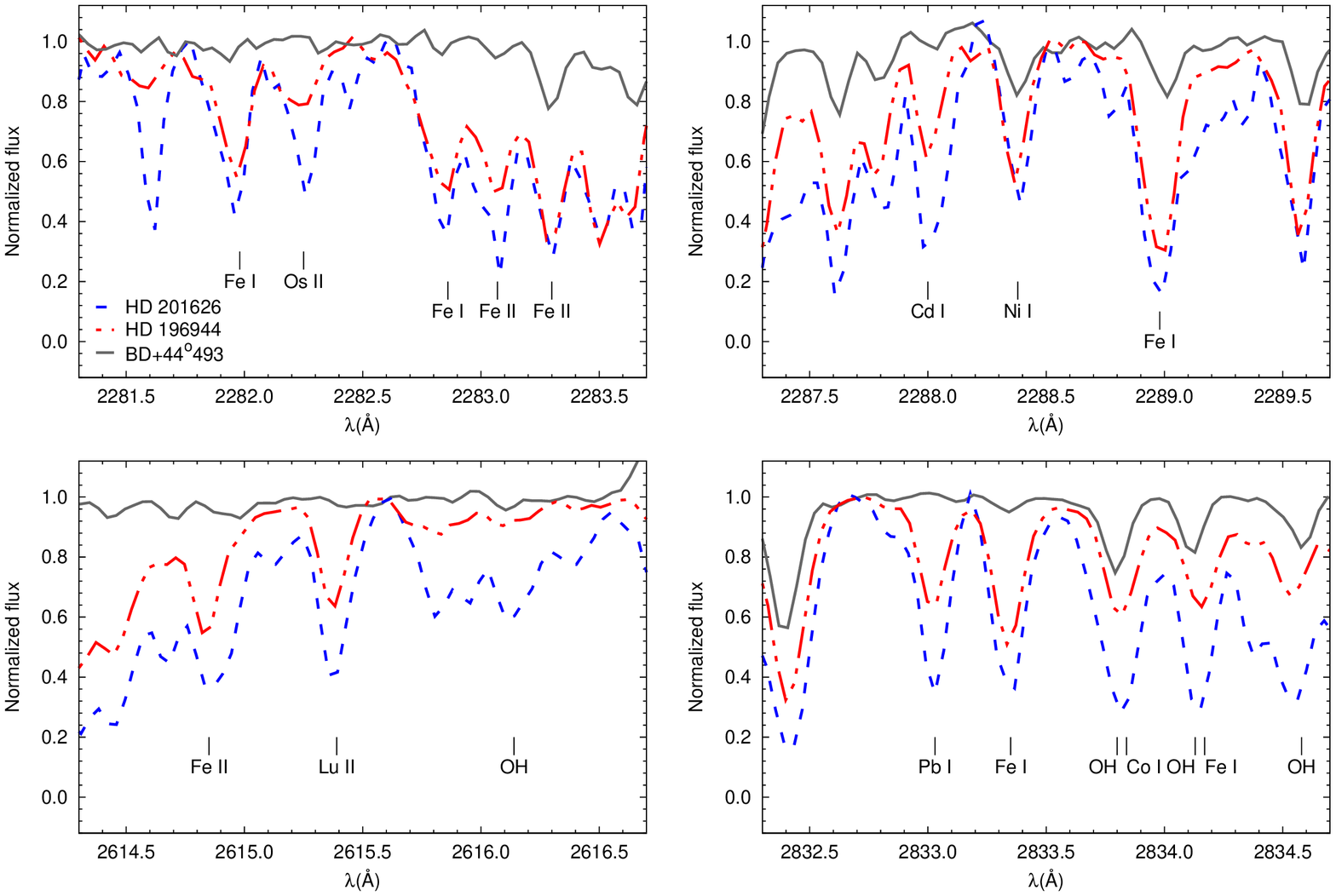}
\caption{\textit{HST}/STIS NUV spectra for \protect\hdb{} and \protect\hda,
compared to the spectrum of \protect\bd, in the regions of the lines of \ion{Os}{2},
\ion{Cd}{1}, \ion{Lu}{2}, and \ion{Pb}{1}.}
\label{nc_comp}
\end{figure}

The NUV spectra of the program stars around the \ion{Mg}{2} doublet at
2800\,{\AA} are shown in Figure~\ref{mg_comp}. For comparison,
\textit{HST}/STIS spectra of the CEMP-no star \bd\ \citep[obtained during this
same program and analysed by][]{placco2014b} is also shown. The effective
temperatures and surface gravities of the program stars are comparable to \bd,
but their metallicities are higher by 1.4$-$2.3~dex.  To better illustrate
the intrinsic differences between these CEMP-$s$ stars and CEMP-no stars,
Figure~\ref{nc_comp} shows portions of the NUV spectra around the lines of four
neutron-capture elements for the same three stars shown in
Figure~\ref{mg_comp}.  Note in particular the absence of absorption by
\ion{Cd}{1}, \ion{Os}{2}, \ion{Lu}{2}, and \ion{Pb}{1} for \bd{}, which signal
a lack of neutron-capture enhancement in \bd{} relative to \hda{} and \hdb.

\subsection{\textit{HIRES} Spectra}

The Keck/HIRES \citep{vogt1994} spectra of both stars were gathered over two
observing runs: 2004 October 1 (program ID: C23H and PI: I.\ Ivans) and 2007
June 5 and 7 (program ID: U100Hb and PI: M.\ Bolte).  \hda{} was observed in
three 300s integrations in both runs.  \hdb{} was observed with three 450s
integrations in program C23H and three 240s integrations in program U100Hb.
The U100Hb data cover the the wavelength range 3725\,{\AA} $\lesssim \lambda \lesssim$
7990\,{\AA}, with resolution $R \equiv \lambda/\Delta\lambda =$ 103,000. 
The C23H data cover bluer wavelengths (2990\,{\AA} $\lesssim \lambda
\lesssim$ 5850\,{\AA}), with resolution $R = 48,000$.

The C23H data were reduced using standard tasks in IRAF\footnote{\footnotesize
IRAF is distributed by NOAO, which is operated by AURA, under cooperative
agreement with the NSF.}, including bias subtraction, bad-pixel interpolation,
and wavelength calibration; and in FIGARO\footnote{\footnotesize FIGARO is
provided by the Starlink Project, which is run by CCLRC on behalf of PPARC
(UK).}, including flat-fielding, light cosmic ray excision, both sky and
scattered-light subtraction, and extraction of the one-dimensional spectra.
The same reduction steps were performed on the U100Hb data using the MAKEE
pipeline (2008 version 5.2.4) \footnote{\href{http://www.astro.caltech.edu/~tb/ipac_staff/tab/makee/index.html}
{http://www.astro.caltech.edu/$\sim$tb/ipac\_staff/tab/makee/index.html}}.
Final processing of the HIRES data, including contiuum normalization and
co-addition of the trios of extracted spectra were done using {\tt
SPECTREdsg} (D.\ S.\ Gregersen, priv.\ comm.), an updated version of the
{\tt SPECTRE} package \citep{SPECTRE}.

\subsection{Atmospheric Parameters}

We determined self-consistent stellar
model-atmosphere parameters using spectroscopic constraints on the
abundances of neutral and ionized species of the same element.  For CEMP and
other cool stars, equivalent-width measurements (EWs) of clean unblended iron
(Fe) lines are usually employed.  However, the NUV region of our stars is
extremely rich and complex.  Thus, we derived our atmospheric parameters
using the EWs of \ion{Fe}{1} and \ion{Fe}{2} lines in the optical
wavelength range.  We then employed this model atmosphere to derive the
remaining abundances at both NUV and optical wavelengths.

Our EWs were obtained using the following tools:

\begin{enumerate}

\item[(i)] Fitting Gaussian profiles to the
observed atomic lines, using the {\tt Robospect} package
\citep{robospect}.

\item[(ii)] Fitting Gaussian and Voigt profiles, using the
{\tt SPECTREdsg} package,
an updated version of the {\tt SPECTRE} package \citep{SPECTRE}.

\item[(iii)] Fitting Gaussian and Voigt profiles, using the {\tt THIMBLES} 
package \citep{THIMBLES}.

\end{enumerate}

We compared the measurements obtained using the {\tt THIMBLES} and {\tt
SPECTREdsg} packages. For 100 lines in common between two sets of independent
measurements, the difference in reduced equivalent widths, REW$\equiv \log(\rm
EW)/\lambda=0.01 \pm 0.05$.
 
The {\tt Robospect} line lists were based on the compilation of
\citet{roederer2012d}, and data retrieved from the VALD database \citep{vald}
and the National Institute of Standards and Technology Atomic Spectra Database
\citep[NIST; ][]{nist}.  The {\tt SPECTREdsg} and {\tt THIMBLES} line lists are
based on those used in \citet{ivans2003, ivans2006}, supplemented and updated
by the work of \citet{bwb2007}, \citet{nilsson2008}, \citet{melendez2009b},
\citet{simmerer2013} \citet{wood2013}, \citet{lawler2014}, \citet{ruffoni+14},
\cite{wood2014a, wood2014b}, and \citet{andersson2015}. 

The atmospheric parameters of the stellar models we employed in our
analysis were derived from these HIRES spectra, employing the following
spectroscopic constraints.   The effective temperature (\teff) was determined
by minimizing any trend between the abundances of \ion{Fe}{1} absorption lines
with the excitation potential ($\chi$); the surface gravity (\logg) was
determined by the equilibrium balance between the abundances of \ion{Fe}{1} and
\ion{Fe}{2}; and the microturbulent velocity ($\xi$) was determined by
minimizing the trend between the \ion{Fe}{1} abundances derived for the
individual lines and their REWs.

For \hda, we derived the best stellar model-atmosphere parameters to be 
5170$\pm$100 / 1.60$\pm$0.25 / $-$2.41$\pm$0.25 / 1.55$\pm$0.10 
[\teff~(K) / \logg~(cgs) / \metal /{} $\xi$~(km~s$^{-1}$)]. 
Our values are consistent with previous determinations from the
literature, listed in Table~\ref{atmlit}.
For \hdb, we derived the best stellar model-atmosphere parameters to be 
5175$\pm$150 / 2.80$\pm$0.45 / $-$1.51$\pm$0.25 / 1.30$\pm$0.10. 
Despite the large literature on \hdb,
the derived stellar model-atmosphere parameters are not in
very good agreement between several different authors, as seen in
Table~\ref{atmlit}.
The \citet{karinkuzhi2014} model-atmosphere parameter estimates agree reasonably
well with ours (within 50~K for \teff{} and 0.1~dex for \metal), apart from a
0.55~dex difference in \logg.  Both studies derived \teff{} values from
spectroscopic constraints on the abundances.  
However, \citet{aoki1997} adopted a \teff{} based on the infrared flux method
\citep{bps1980}, and \citet{sneden2014} adopted one based on the $V-K$ color of
the star \citep{rm2005}, albeit with a low reddening value ($E(V-K) \simeq$
0.171). Table~\ref{atmlit} also contains additional colour-based estimates we
have made of some of the stellar parameters.

To investigate the temperature estimate for \hdb{} issue further, we
performed the following experiment. Using a stellar-atmosphere model
with the \citet{aoki1997} parameter estimates, we find \eps{Fe~I} = 5.50
and \eps{Fe~II} = 5.60; with the \citet{sneden2014} parameters, we find
\eps{Fe~I} = 5.57 and \eps{Fe~II} = 5.30. Thus, demanding a cooler
\teff{} for this star drives the derived metallicity lower, much closer to the
\metal{} adopted by \citet{sneden2014}.   If there was an independent method of
establishing the metallicity of the star other than by the spectroscopic
methods we have employed in this study, a lower \metal{} might indicate that we
have the \teff{} wrong for \hdb. However, the abundances derived from these other 
models do not satisfy the spectroscopic constraints that we employed.

\begin{figure*}[!ht]
\epsscale{1.10}
\plotone{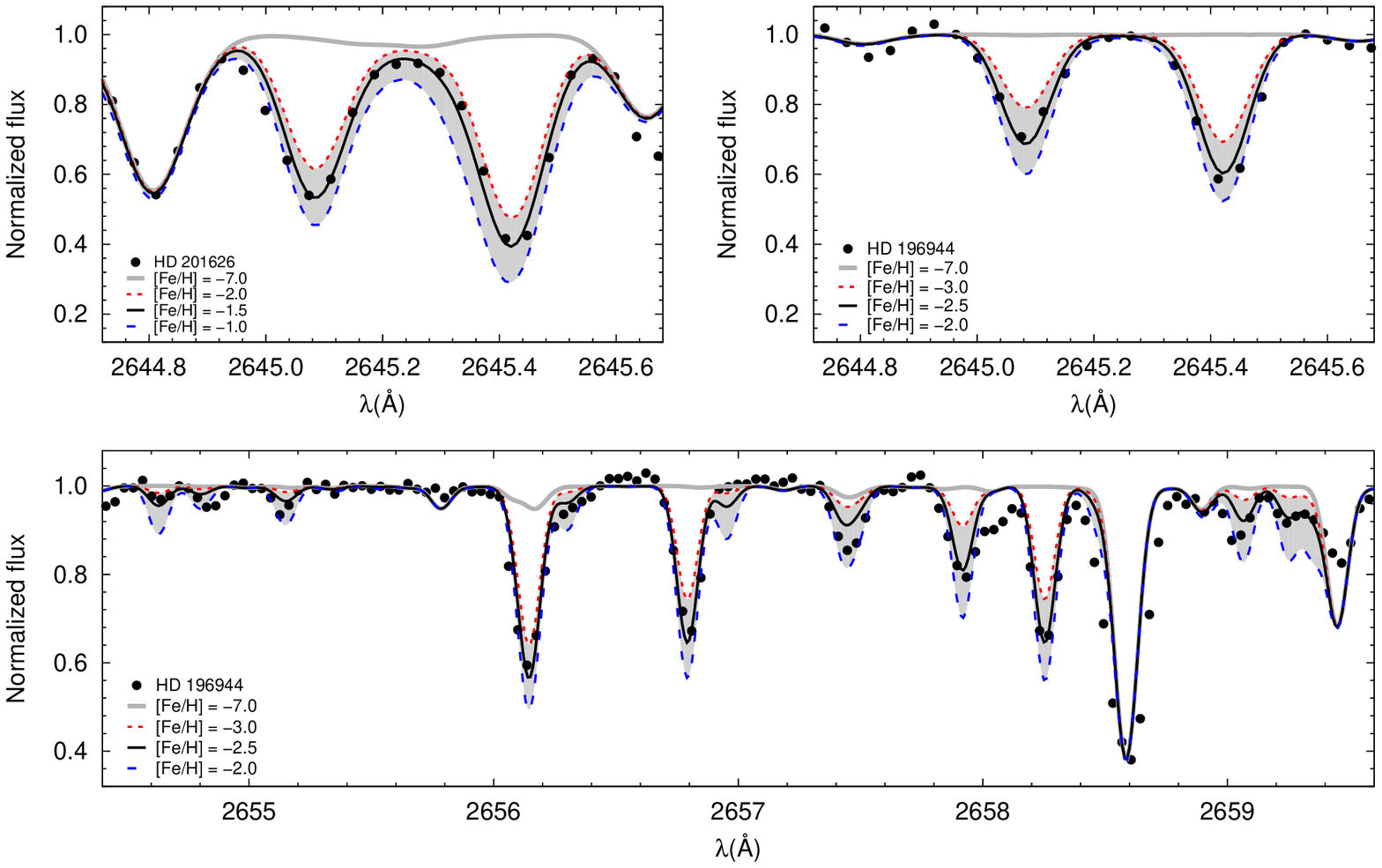}
\caption{Spectral synthesis of NUV \ion{Fe}{1} and \ion{Fe}{2} features for
\protect\hda{} and \protect\hdb. The dots represent the observed spectra, the
solid line is the best abundance fit, and the dotted and dashed line are the
lower and upper abundance limits, used to estimate the abundance uncertainty. The
shaded area encompasses a 1.0~dex difference in \eps{Fe}. The light gray line
shows the synthesized spectrum in the absence of Fe.} 
\label{fesyn} 
\end{figure*}

While we acknowledge the tension between the spectroscopically derived \teff{}
values and those derived over a broader wavelength range for \hdb, the
underlying cause is beyond the scope of this analysis.  Our stellar
model-atmosphere parameters are derived from superb high-resolution
spectra; our line list is extensive (see Table~\ref{abtableOP}); and, as we
show in Section~\ref{absec}, we find excellent agreement between neutral and
ionized states of multiple elements. Of greatest importance for this analysis
is that our spectroscopically derived stellar model-atmosphere parameters
produce spectrum-synthesis calculations that match the observed data
in the NUV wavelengths.

Figure~\ref{fesyn} shows a portion of the NUV spectra of both \hda{} and \hdb,
in a region with a number of \ion{Fe}{1} and \ion{Fe}{2} features.  The dots
represent the observed spectra, the solid line is the best abundance fit,
generated with the atmospheric parameters determined from the optical
spectra.  The dotted and dashed lines limit the shaded area, which encompass a
1.0~dex difference in \metal.  As can be seen from inspection of this figure,
the fit is excellent.

\subsection{Radial-Velocity Variations}

Heliocentric radial velocities for the program stars were determined from clean unblended
lines in the optical spectra. For \hda, we derived 
$V_r = -176.40\pm0.60$~km~s$^{-1}$ on MJD $53279.255346$;
$V_r = -169.40\pm0.30$~km~s$^{-1}$ on MJD $54256.557629$; and
$V_r = -174.30\pm0.20$~km~s$^{-1}$ on MJD $54574.636680$
\footnote{\footnotesize Keck/HIRES data gathered on 2008 April 14 (program ID: U013Hr and
PI: M.\ Bolte), the middle of five exposures.}.  
This star has a number of $V_r$ measurements from the literature: 
$V_r = -174.76\pm0.36$~km~s$^{-1}$ \citep{aoki2002}; 
$V_r = -174.10\pm0.40$~km~s$^{-1}$ \citep{vaneck2003}; 
$V_r = -169.29\pm0.08$~km~s$^{-1}$ \citep{lucatello2005}; 
$V_r = -168.49\pm0.11$~km~s$^{-1}$ \citep{lucatello2005}; 
$V_r = -166.40\pm0.30$~km~s$^{-1}$ \citep{roederer2008}; and
$V_r = -166.80\pm0.70$~km~s$^{-1}$ \citep{roederer2014}.
These radial-velocity variations indicate that \hda{} is a member of a binary
or multiple system. However, there is no published value of its
period based on these measurements. From the values determined by this work and
from the literature, we were able to estimate a period for \hda. We have used a
program described in \citet{buchhave2010}, based on the formalism of
\citet{pal2009}, which models a Keplerian orbit to the data. Assuming initial
guesses of the period and ephemeris, we were able to determine a period of
1325$\pm$12~days, and an eccentricity of 0.015$ \pm$0.036.

For \hdb, we derived 
$V_r = -150.20\pm1.70~km~s^{-1}$ on MJD $53279.325761$; and 
$V_r = -140.00\pm0.30~km~s^{-1}$ on MJD $54258.560998$.  
Literature values include 26 measurements from \citet{mcclure1990}, 
and also the following:
$V_r=-149.40\pm0.80~km~s^{-1}$ \citep{vaneck2003}; 
$V_r=-145.70\pm0.70~km~s^{-1}$ \citep{nordstrom2004}; 
$V_r=-141.60\pm1.20~km~s^{-1}$ \citep{karinkuzhi2014}.
\hdb{} is a confirmed binary with a period of 
1465$\pm$15~days, and an eccentricity of 0.103$ \pm$0.038 \citep{mcclure1990}.
From our derived radial-velocities and the values with available MJD, we were 
able to determine a period of
1470$\pm$9~days, and an eccentricity of 0.090$ \pm$0.051, which are consistent
with the values from the literature.

\begin{figure*}[!ht]
\epsscale{1.10}
\plotone{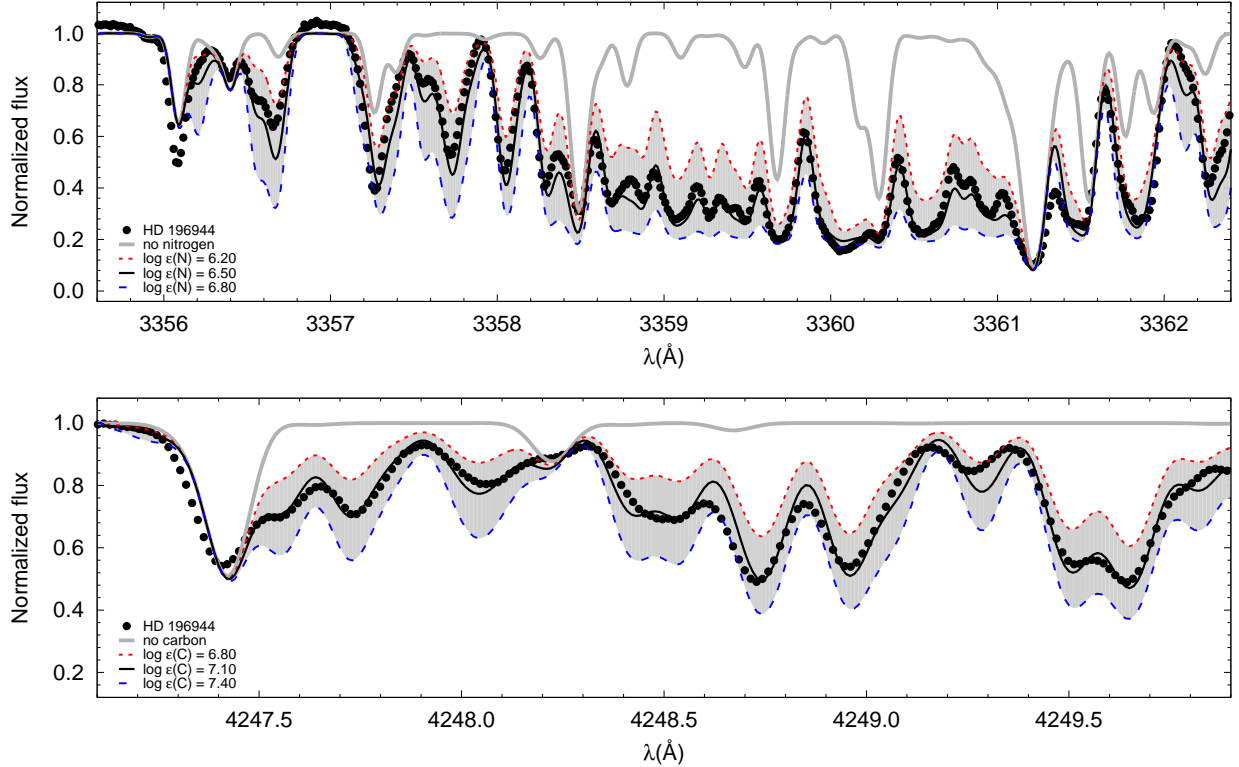}
\caption{Spectral synthesis of the optical NH (top panel) and CH (bottom panel)
features for \protect\hda. The dots represent the observed spectra, the solid
line is the best abundance fit, and the dotted and dashed lines are the lower and
upper abundance limits, used to estimate the abundance uncertainty. The shaded area
encompasses a 0.6~dex difference in \eps{N} and \eps{C}. The light gray line
shows the synthesized spectrum in the absence of N and C.} 
\label{cnsyn}
\end{figure*}

\section{Elemental-Abundance Analysis and Upper Limits}
\label{absec}

Elemental-abundance estimates or upper limits were obtained from the NUV and
optical spectra of \hda{} and \hdb{} for 34 elements: C, nitrogen (N; $Z$=7), O,
sodium (Na; $Z$=11), magnesium (Mg; $Z$=12), silicon (Si; $Z$=14), calcium (Ca;
$Z$=20), scandium (Sc; $Z$=21), Ti, vanadium (V; $Z$=23), Cr, Mn, iron (Fe;
$Z$=26), cobalt (Co; $Z$=27), Ni, zinc (Zn; $Z$=30), Ge, strontium (Sr;
$Z$=38), yttrium (Y; $Z$=39), Zr, Nb, Mo, Cd, barium (Ba; $Z$=56), lanthanum
(La; $Z$=57), cerium (Ce; $Z$=58), neodynium (Nd; $Z$=60), europium (Eu;
$Z$=63), Lu, Hf, Os, Pt, Au, and Pb.

Our abundance analysis utilizes the grid of one-dimensional plane-parallel
ATLAS9 model atmospheres with no overshooting and improved opacity
distribution functions \citep{castelli2004}, computed under the assumption of
local thermodynamic equilibrium (LTE)\footnote{
\href{http://kurucz.harvard.edu/grids.html}
{http://kurucz.harvard.edu/grids.html}}.  We use the 2011 version of the MOOG
synthesis code \citep{sneden1973} for this analysis. To treat isotropic
coherent scattering in this version of MOOG, the solution of the radiative
transfer considers both absorption and scattering components, rather than
treating such scattering as pure absorption \citep[see][for further
details]{sobeck2011}.

Our final abundance ratios, [X/Fe], are given with respect to the solar
abundances of \citet{asplund2009}. Abundances and upper limits for lines
derived from both EWs and spectral synthesis are listed in
Table~\ref{abtableOP} (optical) and Table~~\ref{abtable} (NUV). The average
chemical abundances and upper limits for \hda\ and \hdb\ are listed in
Table~\ref{abfinal2} (optical) and Table~\ref{abfinal} (NUV). The $\sigma$
value refers to the standard error of the mean.  We also evaluated the effect
of changes in each atmospheric parameter on the determined abundances. For this
purpose we used spectral lines with abundances determined by equivalent-width
analysis.  The adopted variations are 150~K for \teff, 0.5~dex for \logg, and
0.3 km\,s$^{-1}$ for $v_\mathrm{micro}$.  Results are shown in Table~\ref{sys}.
Also shown is the total uncertainty, $\sigma_{\rm tot}$, which is calculated
from the quadratic sum of the individual errors.

\begin{figure*}[!ht]
\epsscale{1.10}
\plotone{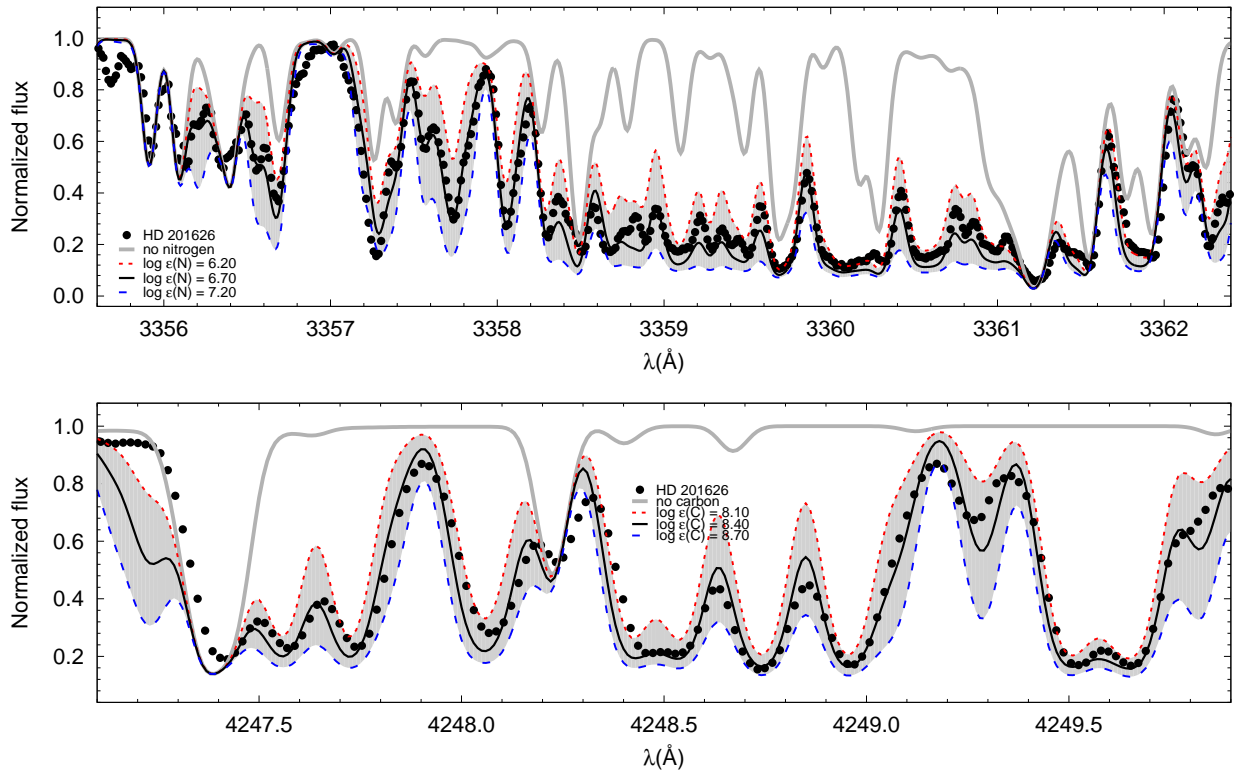}
\caption{Spectral synthesis of the optical NH (top panel) and CH (bottom panel)
features for \protect\hdb. The dots represent the observed spectra, the solid
line is the best abundance fit, and the dotted and dashed line are the lower and
upper abundance limits, used to estimate the abundance uncertainty. The shaded area
encompasses a 1.0~dex difference in \eps{N} and 0.6~dex in \eps{C}. The light
gray line shows the synthesized spectrum in the absence of N and C.} 
\label{cnsyn2}
\end{figure*}

\subsection{Carbon, Nitrogen, and  Oxygen}

Carbon, nitrogen, and oxygen abundances were determined for both stars,
using the NUV and optical spectra, from both EW analyses and
spectral synthesis calculations.

\begin{figure*}[!ht]
\epsscale{1.10}
\plotone{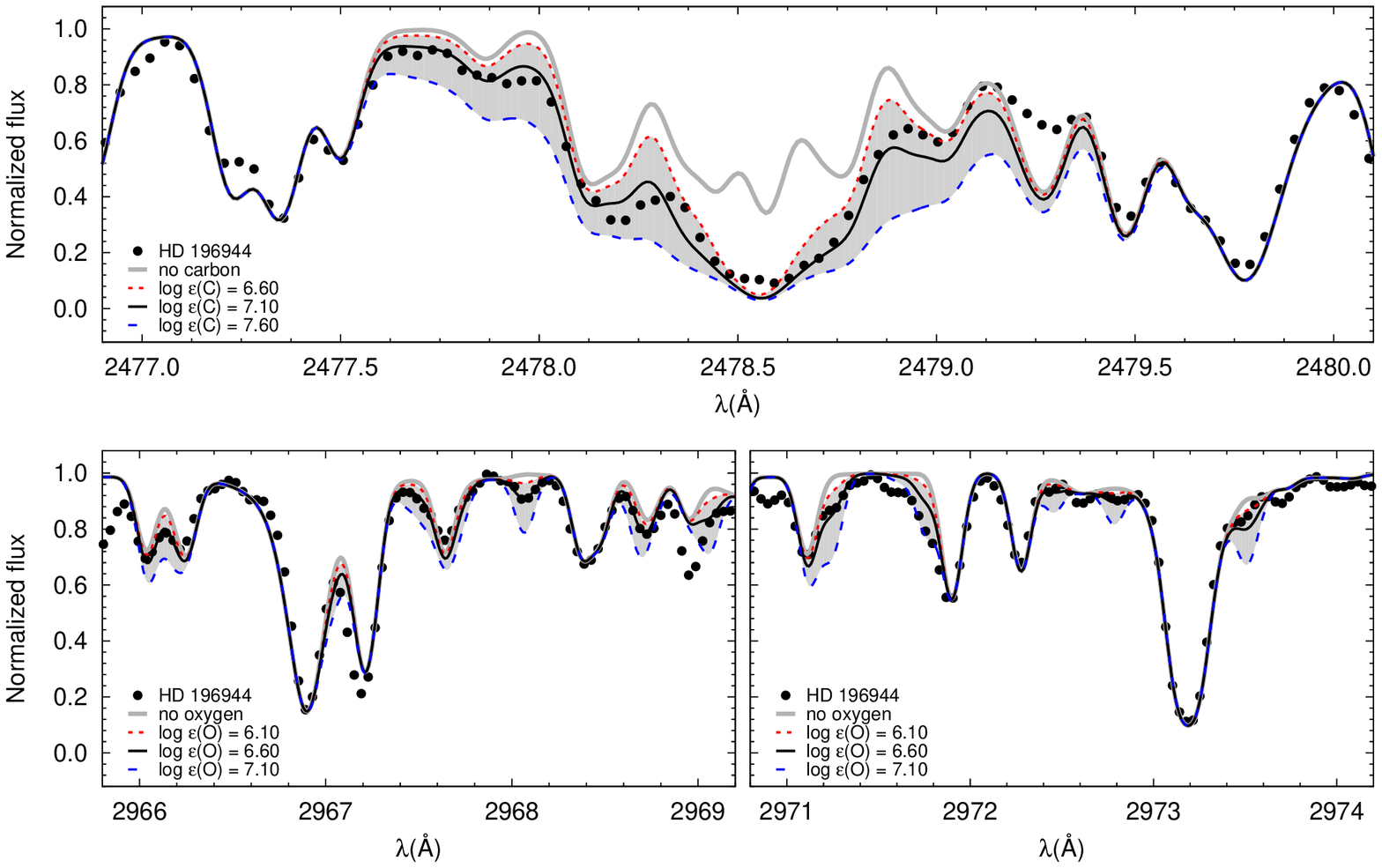}
\caption{Spectral synthesis of the NUV \ion{C}{1} (top panel) and OH (bottom panel)
features for \protect\hda. The dots represent the observed spectra, the solid
line is the best abundance fit, and the dotted and dashed line are the lower and
upper abundance limits, used to estimate the abundance uncertainty. The shaded area
encompasses a 1.0~dex difference in \eps{C} and \eps{O}. The light gray line
shows the synthesized spectrum in the absence of C and O.} 
\label{cosyn2}
\end{figure*}

Carbon abundances were determined from optical CH features at 4246\, {\AA} and
4313\,{\AA}. For \hda, both regions are consistent with \eps{C}=$7.10$ and,
for \hdb, \eps{C}=$8.40$. The lower panels of Figures~\ref{cnsyn} and
\ref{cnsyn2} show the synthesis for the 4246\, {\AA} region in both stars.
We derived an isotopic ratio $^{12}$C/$^{13}$C = 4 for \hda, and
$^{12}$C/$^{13}$C = 50 for \hdb. The high $^{12}$C/$^{13}$C for \hdb{} is
consistent with AGB models, and the low value for \hda{} can be explained
by its evolutionary stage, in which internal mixing slightly decreased the
$^{12}$C/$^{13}$C ratio. We also attempted to determine the carbon
abundance from the \ion{C}{1} atomic features at 2478\,{\AA} and 2967\,
{\AA}. For \hda, the \eps{C}=$7.10$ value determined for the optical
spectra can be well-reproduced for the 2478\,{\AA} line (upper panel of
Figure~\ref{cosyn2}).  The agreement between abundances determined from the
atomic (NUV) and molecular (optical) features is worth noting, since
previous works \citep{asplund2005,collet2007} have suggested that 3D
effects on the CH and C$_2$ bands could lead to overestimates of [C/Fe] of
$+0.5$ to $+0.8$ dex in metal-poor subgiants.  For \hdb, both regions are
heavily blended with other species, and it was not possible to determine
abundances. We also calculated the carbon-abundance corrections for both
stars, based on the procedure described by \citet{placco2014c}.  This
procedure takes into account the evolutionary status of the star (from the
\logg{} value) to find a correction for the carbon depletion, which occurs
during stellar evolution on the giant branch. Since \hda{} is a red
horizontal-branch star, we take its carbon correction only as a lower
limit. Using the [N/Fe]=0.0 case, we find corrections of $>$ +0.23~dex for
\hda, and 0.02~dex for \hdb.

Nitrogen abundances were determined from the NH feature at 3360\,{\AA}.  
The line list was provided by \citet{kurucz}, following the procedure
described in \citet{aoki2006}.
For
\hda, the best value is \eps{N}=$6.50$ (upper panel of Figure~\ref{cnsyn})
and, for \hdb, \eps{N}=$6.70$ (upper panel of Figure~\ref{cnsyn2}).

Oxygen abundances were determined from NUV spectral synthesis of OH features
for \hda, and spectrum synthesis of the [\ion{O}{1}] 6301\,{\AA} line for \hdb.
For \hdb, we were also able to measure the O triplet near 7700\,{\AA}.
However, we do not include the abundances from these lines in our analysis,
for reasons we describe below.

The O-triplet features possess high excitation potentials, and are expected to
suffer from non-LTE effects. For \hdb, LTE analysis of the O-triplet EWs yields
\eps{O}$=8.15$, with a standard deviation of 0.17 dex (based on the
line-to-line scatter).  Applying the recommended correction of
\citet{takeda2003} for each of the three lines, we derive \eps{O}$=8.05$, which
is closer to that of the abundance we derive from spectrum-synthesis
calculations of the [\ion{O}{1}] 6301\,{\AA} feature.  A star with atmospheric
parameters close to those of \hdb{} was studied by \citet{garciaperez2006}
(HD~274949: 5090/2.76/$-$1.51), who independently calculated non-LTE abundance
corrections for the triplet.  Their $\Delta$\eps{O} correction is $-0.10$ dex,
in agreement with the value we find adopting the recommended correction of
\citet{takeda2003}.  However, in a comparison against \citet{takeda2003},
\citet{fabian+2009} find the necessary non-LTE corrections to be significantly
larger than those reported by \citet{takeda2003} for stars warmer than 5500~K.
\citet{fabian+2009} also note that the stellar metallicity plays an important
role in the degree of the non-LTE effect.  For the stars included in the study
of \citet{ramirez+2013} that are closest in their stellar parameters to \hdb{}
(HIP~60719: 5250/2.7/$-$2.42 and HIP~18235: 5010/3.2/$-$0.73) the corrections
to the LTE \eps{O} abundance derived from the triplet are $\Delta$\eps{O} of
+0.66~dex and $-$0.29~dex, an order of magnitude difference.  Thus, in our
analysis of \hdb, we rely on the abundance derived from the [\ion{O}{1}]
6301\,{\AA} line.  

There are several OH features in the 2965--2975\,{\AA}
region of \hda{} (lower panels of Figure~\ref{cosyn2}). These were all
be reasonably fit with a \eps{O}=$6.60$. However, we conservatively
adopt a 0.3~dex uncertainty, due to the presence of a few unknown
features in this region. 

\subsection{The Iron-Peak Elements}

\begin{figure}[!ht]
\epsscale{1.20}
\plotone{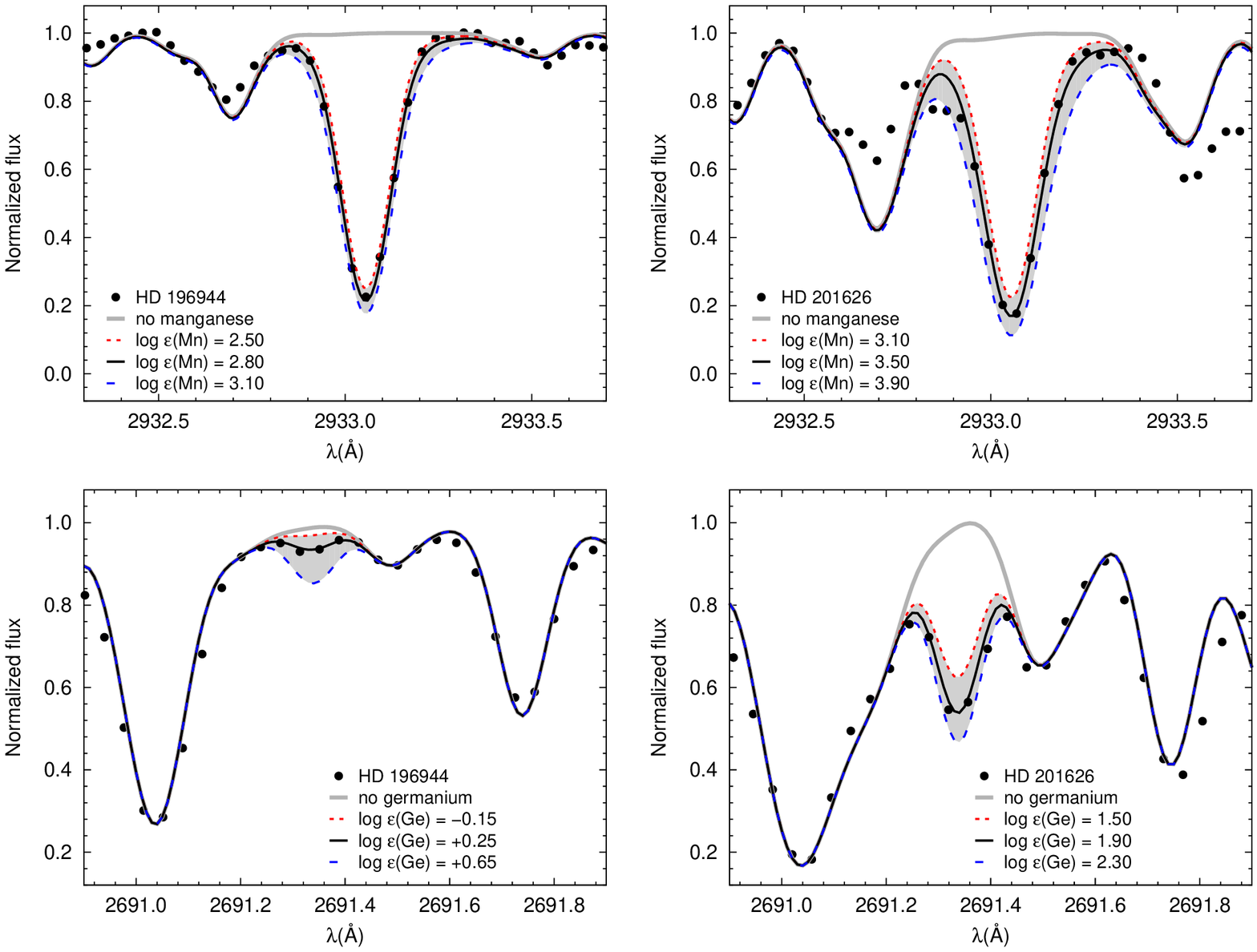}
\caption{Spectral synthesis of the NUV \ion{Mn}{2} and \ion{Ge}{1} features for
\protect\hda{} and \protect\hdb. The dots represent the observed spectra, the
solid line is the best abundance fit, and the dotted and dashed line are the
lower and upper abundance limits, used to estimate the abundance uncertainty. The
shaded area encompasses a 0.6--0.8~dex difference in \eps{Mn} and \eps{Ge}. 
The light gray line shows the synthesized spectrum in the absence of Mn and Ge.} 
\label{mngesyn}
\end{figure}

Abundances for Ti, Cr, and Ni from the NUV spectra were determined for \hda\
with an EW analysis. No such measurements were made for \hdb, since it is more
metal-rich and the same NUV lines are heavily blended. We derived \ion{Mn}{2}
abundances from spectral synthesis of NUV features at 2933\,{\AA} and
2949\,{\AA} for both stars, because these lines are broadened by hyperfine
splitting of the $^{55}$Mn isotope.  Figure~\ref{mngesyn} (upper panels) shows
the synthesis of the 2933\, {\AA} \ion{Mn}{2} line in the NUV spectrum for both
program stars. We were also able to determine the abundance of Ge in both
stars.  This element belongs to the transition between iron-group and
neutron-capture elements. Three lines (2651\,{\AA}, 2691\,{\AA}, and 3039\,{\AA}) were
measured for \hda, and one line (2691\,{\AA}) for \hdb. The lower panels of
Figure~\ref{mngesyn} show the spectral synthesis for the \ion{Ge}{1}
2691\,{\AA} line. 

A number of iron-peak element abundances were determined from EW analysis of the
optical spectra of \hda{} and \hdb.  Comparing the optical and NUV
determinations, we find good agreement for Ti, Cr, and Ni for \hda{},
and Mn for both stars. For elements with two different ionization states
measured in \hda, abundances are within 0.04~dex for Ti and 0.22~dex for Cr
obtained from the optical data, and 0.02~dex for Ti,  0.18~dex for Cr, and
0.40~dex for Ni from the NUV data. The good agreement between the \ion{Ti}{1}
and \ion{Ti}{2} also confirms our spectroscopically determined \logg{} values.
The large difference for Ni lines in the NUV is mainly due to the difficulty of
measuring EWs for the \ion{Ni}{2} lines.

\subsection{The Neutron-Capture Elements}

For the neutron-capture elements (Cu to Pb), all of the abundances listed in
Tables~\ref{abfinal2} and \ref{abfinal} were determined from spectrum-synthesis
calculations, using both optical and NUV spectra. For the optical spectra, some
elements had their abundances first estimated from their EW that were used as
starting points for the synthesis.

\begin{figure}[!ht]
\epsscale{1.20}
\plotone{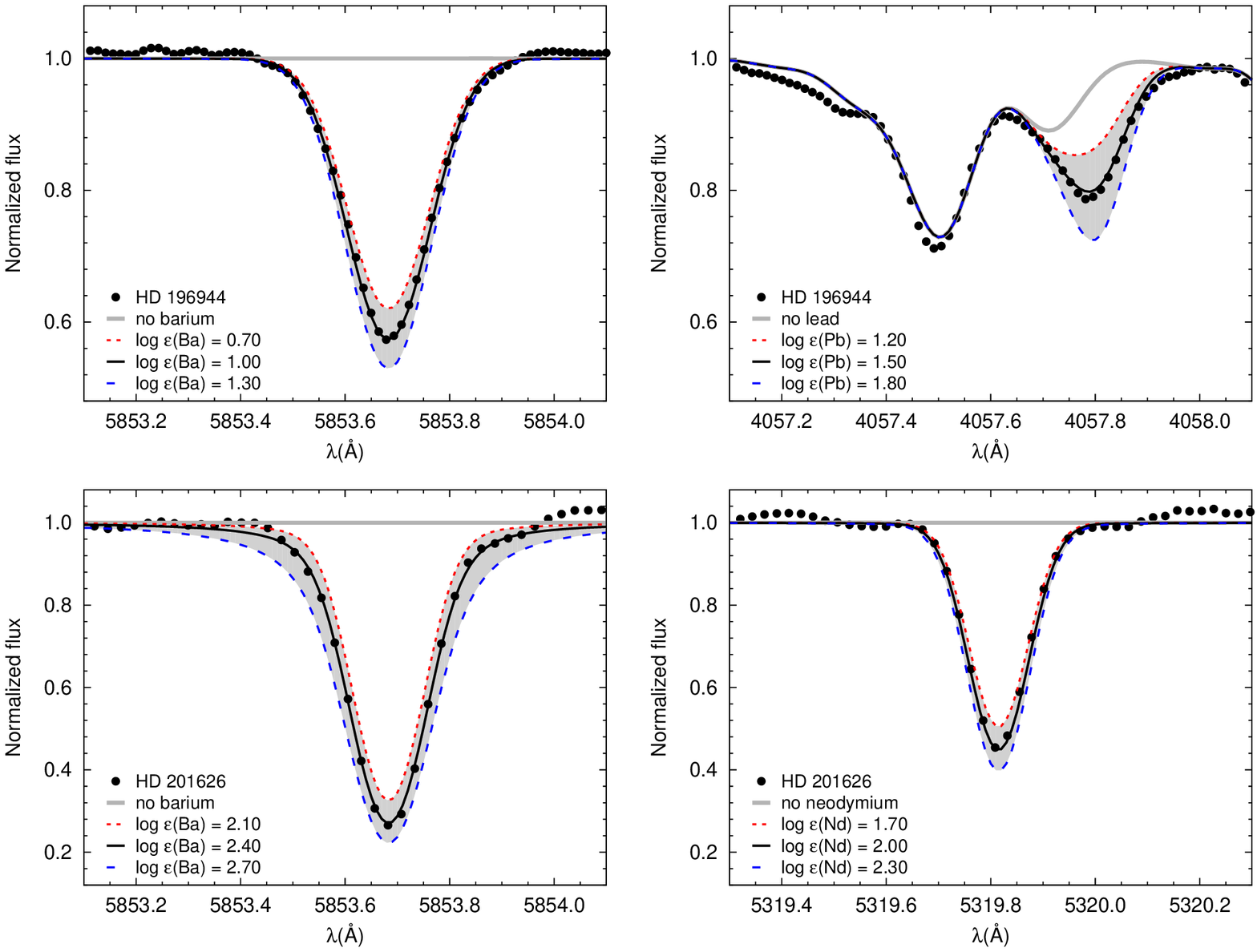}
\caption{Spectral synthesis of the optical \ion{Ba}{2}, \ion{Nd}{2}, and \ion{Pb}{1}
features for \protect\hda\ and \protect\hdb. The dots represent the observed
spectra, the solid line is the best abundance fit, and the dotted and dashed
line are the lower and upper abundance limits, used to estimate the abundance
uncertainty. The shaded area encompasses a 0.6~dex difference in the abundances.
The light gray line shows the synthesized spectrum in the absence of the listed
species.} 
\label{optsyn}
\end{figure}

There are a number of unblended \ion{Zr}{2} lines in the NUV region of the
spectra of these stars. For both \hda{} and \hdb, the agreement between the
individual abundances is generally good (between 0.1--0.2~dex), apart from the
2699\,{\AA} feature, which is consistently lower by about 0.5~dex.  \ion{Ba}{2}
lines are broadened by hyperfine splitting (hfs) of the $^{135}$Ba and
$^{137}$Ba isotopes; so even though we also determined Ba abundances from EW
analysis, the final average values come from spectral synthesis only.
Figure~\ref{optsyn} shows the optical spectral synthesis of the \ion{Ba}{2}
5853\,{\AA} line for both stars, the \ion{Pb}{1} 4057\,{\AA} line for \hda, and
the \ion{Nd}{2} 5319\,{\AA} line for \hdb. The EW Ba abundances for \hda{}
agree within 0.1~dex with the spectral synthesis for the 4554\,{\AA},
5853\,{\AA}, and 6141\,{\AA} lines (\eps{Ba}=1.0).  For \hdb, we report the
abundance from the spectrum synthesis of the 5853\,{\AA} feature.  The
synthesis of the 6141\,{\AA} and 6406\,{\AA} features are consistent with this
value, but the 4554\,{\AA} region is too complex to model
well\footnote{\footnotesize Using the \teff=4800~K models, spectrum-synthesis
calculations of molecule-rich regions showed no improvement over the syntheses
produced employing the model derived in this study.}.

\begin{figure}[!ht]
\epsscale{1.20}
\plotone{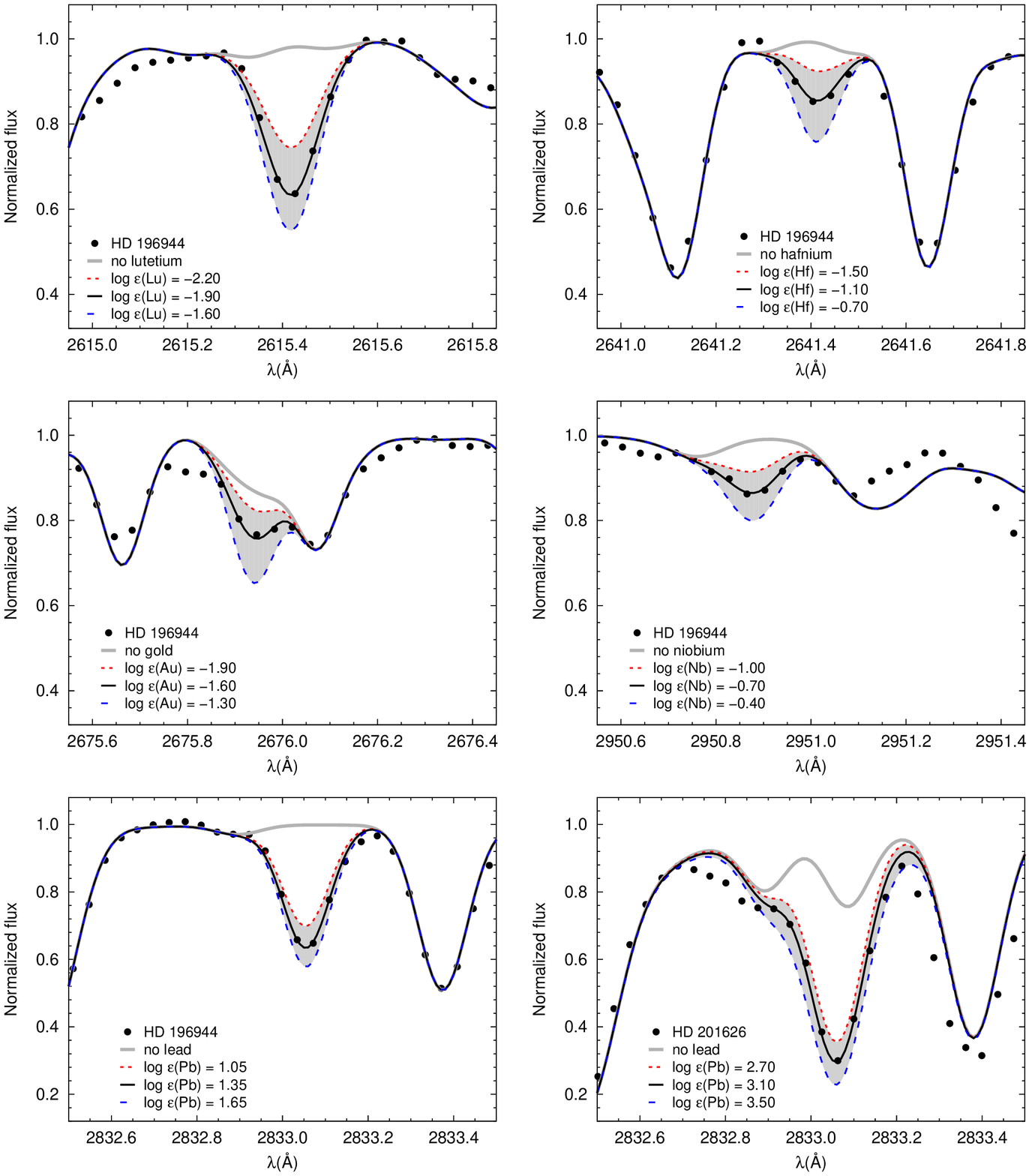}
\caption{Spectral synthesis of the NUV \ion{Lu}{2}, \ion{Hf}{2}, \ion{Au}{1}, and
\ion{Nb}{2} features for \protect\hda, and \ion{Pb}{1} for both stars. The dots
represent the observed spectra, the solid line is the best abundance fit, and
the dotted and dashed line are the lower and upper abundance limits,
used to estimate
the abundance uncertainty. The shaded area encompasses a 0.6--0.8~dex difference
in the abundances. The light gray line shows the synthesized spectrum in the
absence of the listed species.} 
\label{nuvsyn}
\end{figure}

Figure~\ref{nuvsyn} shows the spectral synthesis of Lu, Hf, Au, and Nb for
\hda, and Pb (2833\,{\AA} line) for both stars.  Pb abundances determined from
neutral species are more affected by non-LTE effects than abundances determined
from ionized species.  For the stellar parameters of the stars studied in this
work, this effect can increase the Pb abundance by up to 0.5~dex
\citep{mashonkina2012}. This effect is somewhat counterbalanced by considering
3D modeling \citep{siqueira2013}. For simplicity, we consider the combination
of both effects to be within the uncertainties of the spectral synthesis.
Overall, we were able to reproduce the observed spectra with our line list,
even with difficult blends, such as the one for the \ion{Au}{1} 2675\,{\AA}
line.  
For \hda, there is a small discrepancy between the Pb abundances determined
from the 2833\,{\AA} (\eps{Pb}=1.35) and 4057\,{\AA} (\eps{Pb}=1.50) lines. 

\begin{figure*}[!ht]
\epsscale{1.10}
\plotone{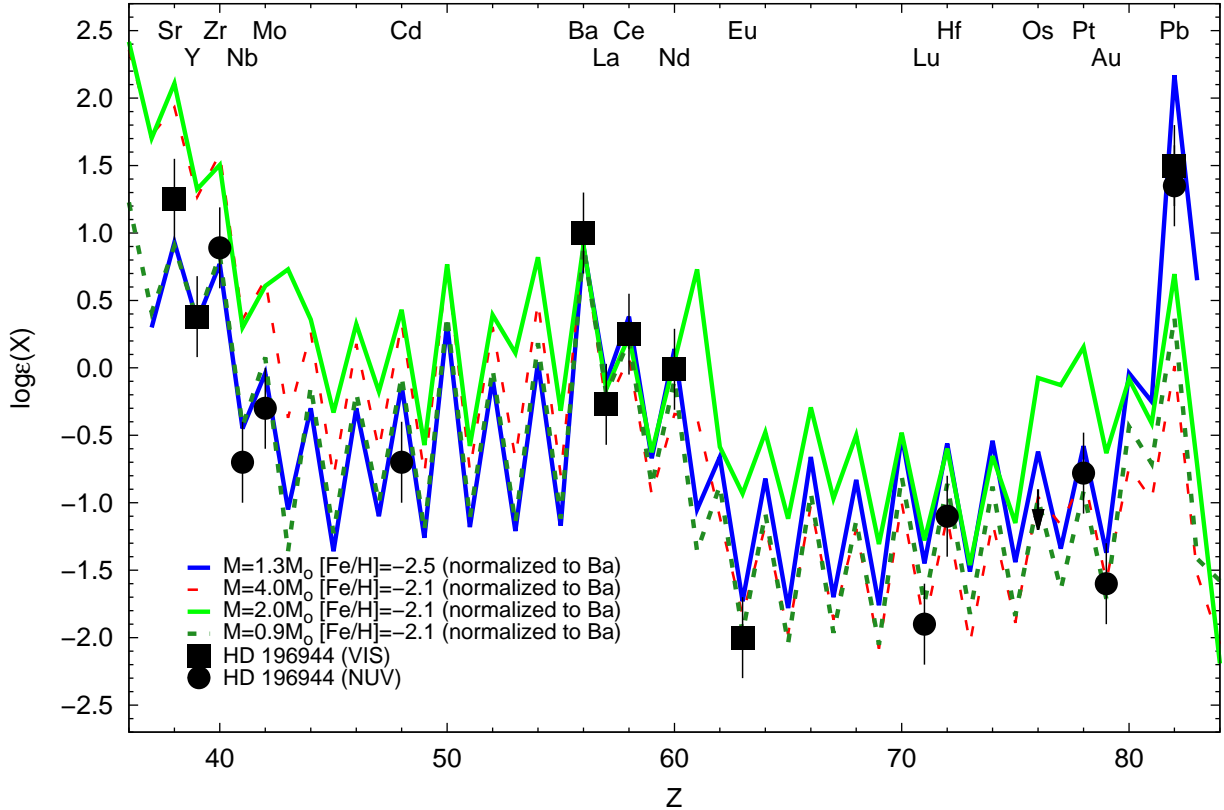}
\caption{Comparison between elemental-abundances and upper limits determined from this
work for \protect\hda, and theoretical yields for four different AGB
models from \citet{bisterzo2011} and \citet{placco2013}. 
The models are normalized to the optical Ba abundance estimate.}
\label{patternba}
\end{figure*}

\section{Discussion}
\label{diss}

The general behavior of the elemental-abundance patterns for CEMP-$s$ stars is
understood and well-explained by the evolution in a binary system with a more
massive companion going through the AGB phase and polluting the atmosphere of
the currently observed, less massive star. However, there is still debate on
the range of masses of the AGB donor star, as well as the details of the mass
transfer, dilution effects, and orbital features of the binary system.
Elemental and isotopic abundances for CEMP-$s$ stars can help distinguish between different
models. For \hda{} and \hdb, we evaluated the chemical abundance patterns by
comparing their optical and NUV determinations with AGB yields from
\citet{bisterzo2010} and \citet{placco2013}, as well as a model of binary
evolution and AGB nucleosynthesis from \citet{abate2015a}.

\subsection{Comparison with Bisterzo et al.\ (2010)}

Figure~\ref{patternba} shows the optical and NUV abundances for \hda,
compared with AGB yields from the model presented in \citet{placco2013},
with M = 1.3~M$_{\odot}$ and \metal=$-$2.5, as well as three additional
models from \citet{bisterzo2010}, with \metal=$-$2.1 and M = 0.9, 2.0,
and 4.0~M$_{\odot}$. Although all the models well-reproduce the
second peak of the $s$-process (Ba-Nd), the more massive AGB stars
over-produce the first-peak $s$-process elements (Sr, Y, and Zr). In
addition, Pb is under-produced by the \metal=$-2.1$ models, and the
observed values are closer to the \metal=$-2.5$ model. \citet{bisterzo2011} also
studied \hda, and found the best match with a model of M = 1.5~M$_{\odot}$, with
no initial enhancements of $r$-process elements.

\begin{figure}[!ht]
\epsscale{1.20}
\plotone{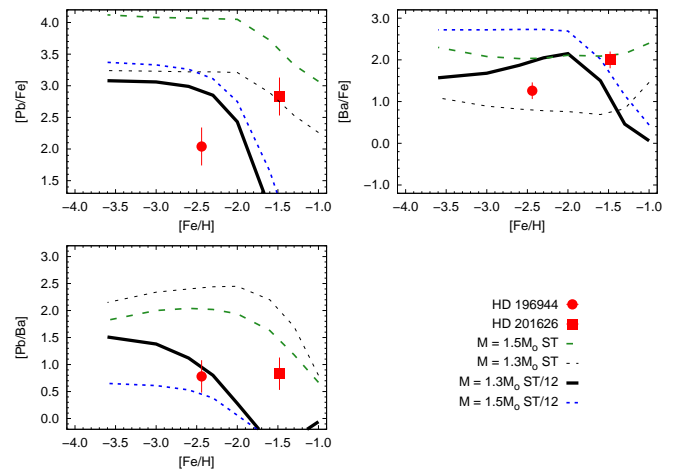}
\caption{\abund{Pb}{Fe}, \abund{Ba}{Fe}, and \abund{Pb}{Ba} compared with AGB 
theoretical predictions for two different initial masses and $^{13}$C-pocket 
efficiencies. Data were taken from Tables B5 and B6 of \citet{bisterzo2010}.}
\label{bisterzo}
\end{figure}

The \citet{bisterzo2010} models also calculate the evolution of $s$-process element
abundances with \metal, which is a useful diagnostics for selecting AGB models
with different masses and $^{13}$C-pocket efficiencies.  Figure~\ref{bisterzo}
shows the \abund{Pb}{Fe}, \abund{Ba}{Fe}, and \abund{Pb}{Ba} values for \hda{}
and \hdb{} compared with four of the model prescriptions from
\citet{bisterzo2010}.  The [Pb/Ba] values for \hda{} and \hdb{} are consistent
with the $s$-process nucleosynthesis in low-mass low-metallicity AGB stars with
the highest neutron exposures possible.

\begin{figure*}[!ht]
\epsscale{1.20}
\plotone{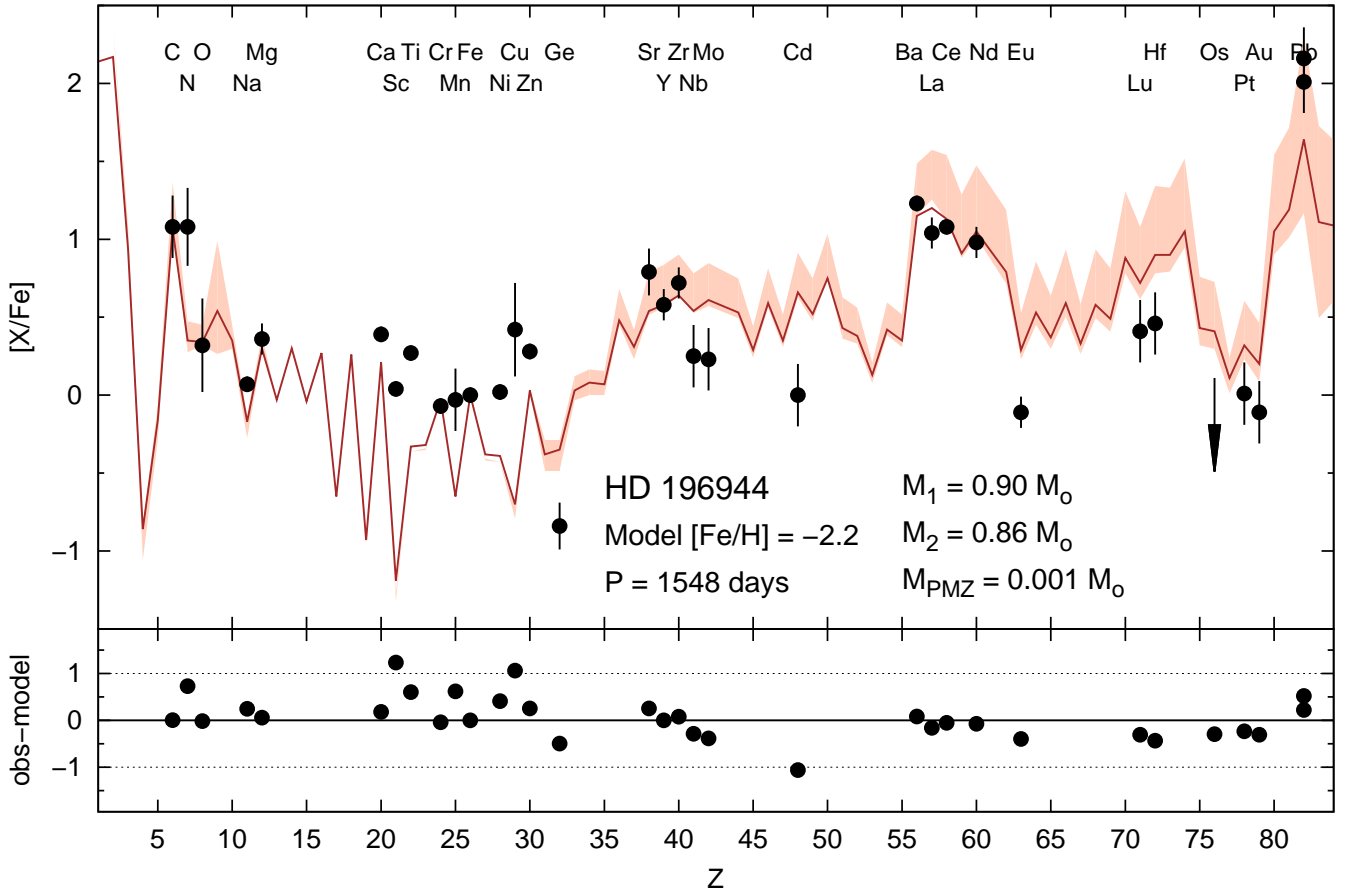}
\caption{Upper panel: Model fitting for \protect\hda. The black filled dots are
abundances determined by this work, the solid line is the best-fitting model,
and the shaded area represents the uncertainty on the model values (see
Section~\ref{abatesec} for details). Lower panel: Residuals computed as the
difference between the observed and model abundances.}
\label{abateHD19}
\end{figure*}

\citet{bisterzo2010} also point to the importance of measuring Nb abundances in
CEMP stars, to distinguish between the $intrinsic$ (star on the
thermally-pulsing AGB or post-AGB phase) and $extrinsic$ (main-sequence or red
giant-branch star -- CEMP-$s$ stars).  
As noted by \citet{wd1988}, an $intrinsic$ AGB star is expected to be
technetium-rich (Tc; $Z$=43), $^{93}$Zr-rich, and $^{93}$Nb-poor, with
\abund{Zr}{Nb}$\sim$1.  The $s$-process abundances of an $extrinsic$ AGB star
are the result of pollution from a former AGB star onto the atmosphere of what
once was the lower mass star in a binary system where the abundance of Nb
(which has a single stable isotope) is a result of $\beta$-decay from
$^{93}$Zr, leading to \abund{Zr}{Nb}$\sim$0 \citep[see e.g. the CEMP-$r/s$ star
CS~29497$-$030, with \abund{Zr}{Nb}$= -$0.27,][]{ivans+2005}.  For \hda, for which these
two elemental abundances were derived from the NUV spectra, we find
\abund{Zr}{Nb}=$+0.47 \pm 0.21$. However, we note that our Nb abundance is
derived from the weak \ion{Nb}{2} feature at 2950\, {\AA}.

\subsection{Comparison with Abate et al.\ (2015a)}
\label{abatesec}

We also investigated the properties of \hda{} and \hdb{} by comparing their
derived elemental abundances with the grid of binary-evolution models of
\citet{abate2015a}. This grid consists of about $290,\!000$ binary stars
distributed in the $M_1-M_2-\log_{10}a-M_{\mathrm{PMZ}}$ parameter space, where
$M_{1,2}$ are the initial masses of the primary and secondary stars,
respectively, $a$ is the initial orbital separation of the system, and
$M_{\mathrm{PMZ}}$ is the mass of the partial-mixing zone.
The mass of the partial-mixing zone is a parameter
of the nucleosynthesis model that relates to the amount of free neutrons
available for neutron-capture processes in AGB stars \footnote{For details
about the modelling of the partial-mixing zone and its role in AGB
nucleosynthesis, we refer the interested reader to
\citet{karakas2010,lugaro2012, abate2015a}.}. In this study we used model set B
of \citet{abate2015a}, which allows efficient angular-momentum loss during mass
transfer and large wind mass-accretion efficiencies over a wide range of
orbital separations.

The best-fitting model to the observed abundances of each star is determined
following \citet{abate2015a}. The model ages are selected between 10 Gyr and
13.7 Gyr, bracketing the expected ages of very metal-poor halo stars. The
orbital periods of the models are chosen to reproduce the observed periods
within the spatial resolution of our grid ($\Delta\log_{10}(a/R_{\odot})=0.1$,
which corresponds to $\Delta\log_{10}(P_{\mathrm{orb}}/\mathrm{days})=0.15$).
To constrain the evolutionary stage, we select model stars that reproduce the
measured surface gravities within the observational uncertainties. For the
model stars that fulfill these criteria, we determine the model that minimizes
$\chi^2$, calculated by taking into account the abundances of the light
elements up to atomic number 12 (Mg), and all neutron-capture elements from Ga
($Z$=31) to Pb ($Z$=82). The elements with atomic numbers between 13 (Al) and
30 (Zn) are not considered in the calculation of $\chi^2$, because they are not
in general produced by low-mass stars \citep[$M < 3 M_{\odot}$, see,
e.g.,][]{karakas2009,cristallo2011}. 
In cases where only upper limits are available for a given element, an observational
uncertainty of 0.3 dex is adopted to compute $\chi^2$, if the model abundance value is 
higher than the observed upper limit. 
This value was chosen because it is the largest observational
uncertainty from our abundance determinations.

The initial abundances used in the models (up to Ge) are predicted by
\citet{kobayashi2011c} in their models of Galactic chemical evolution at
\metal=$-$2.3. For elements heavier than Ge, we use Solar System scaled values.
It must be kept in mind that the impact of the choice of the initial set of
abundances is negligible when studying CEMP-$s$ stars. This is because the
abundances of most key elements (e.g., C, N, Na, Mg, and all neutron-capture
elements) have large variations during AGB evolution.
The fit is made by combining the abundances of C and N. This is because, in
giants, some amount of C is converted to N at the bottom of the
convective envelope during dredge up. Even though the exact amount is
very uncertain, the total C$+$N is conserved, therefore the total C$+$N
predicted by the models is supposed to reproduce the observed abundance.
For this comparison, we thus do not employ the carbon-abundance
corrections, just the observed values.

\begin{figure*}[!ht]
\epsscale{0.90}
\plotone{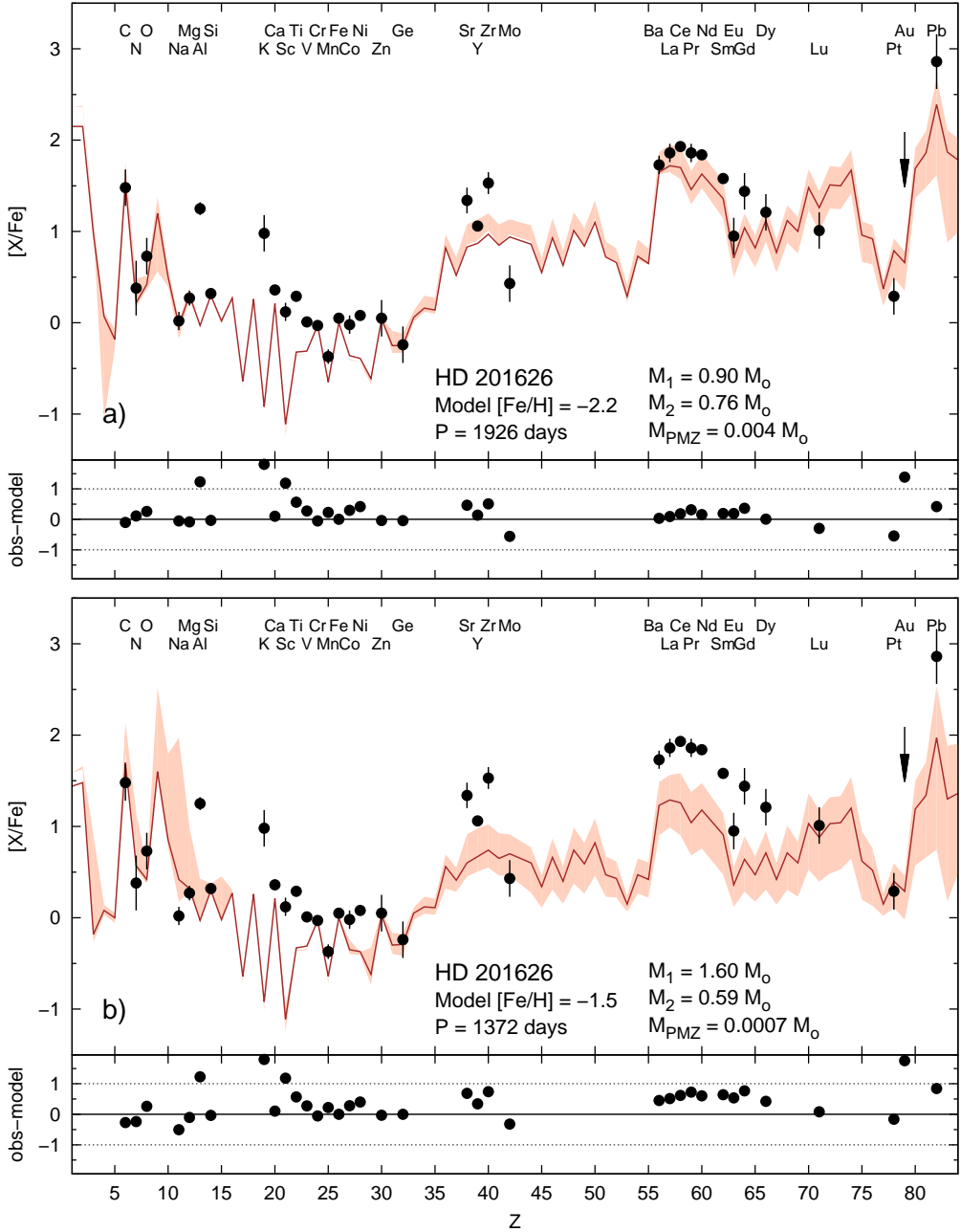}
\caption{Same as Figure~\ref{abateHD19} and: a) with model metallicity of
$Z = 10^{-4}$ (corresponding to [Fe/H]$\approx-2.2$). b) with model metallicity of
$Z = 5\cdot10^{-4}$ ([Fe/H]$\approx-1.5$).}
\label{abateHD20}
\end{figure*}

Figures~\ref{abateHD19} and \ref{abateHD20} show the best-fitting models for
both \hda{} ($M_{1,i}=0.9 M_{\odot}$, $M_{2,i}=0.86 M_{\odot}$, for \metal=$-2.2$),
and  \hdb{} ($M_{1,i}=0.9 M_{\odot}$, $M_{2,i}=0.76 M_{\odot}$, for \metal=$-2.2$;
$M_{1,i}=1.6 M_{\odot}$, $M_{2,i}=0.59 M_{\odot}$, for \metal=$-1.5$)
 as solid lines. 
The abundances from this work are shown
as filled dots with error bars.  The shaded areas encompass the minimum and
maximum abundances predicted by models that: (i) have an age between 10 Gyr and
13.7 Gyr; (ii) reproduce the observed period within the grid resolution and
\logg within the observational uncertainties; and (iii) have a final $\chi^2$
that is less than three times the $\chi^2$ of the best fit.  For \hdb, we ran
two models, with metallicities $Z$ = $10^{-4}$ and $5\times10^{-4}$, which
correspond to \metal=$-$2.2 and $-$1.5, respectively, with the solar abundance
set of \citet{asplund2009}.  Also shown in the plot are the initial parameters
of the best models (\metal, M$_1$, M$_2$, M$_{\rm PMZ}$) and the final orbital
period (given as input).  Comments on the model-matching for each star are
given below.

\subsubsection{HD~196944}

\hda{} is a CEMP-$s$ (\abund{Ba}{Eu}=$+$1.33) giant star with \metal = $-2.46$.
This star has been the subject of many studies in the literature -- SIMBAD
notes 67 references.  This star is known to be an extremely weak-lined G/K star
with very strong CH bands \citep{bidelman1981}. For the model comparison, we
used the abundances from both optical and NUV spectra, as well as the orbital
period determined in this work.  To test the robustness of the model-fitting
procedure, we made a few experiments, removing abundances for elements mainly
formed by the $r$-process (e.g., Eu, Lu, Os, Pt, and Au); the initial
parameters adjusted with the routines did not change.

Inspection of Figure~\ref{abateHD19} reveals excellent agreement between
the observations and model predictions for the light elements C, O, Na, Mg, Cr, and
Fe. Some unexpected differences are found for N, Sc, and Ti. For germanium, even
though the abundance is $\sim$0.5~dex lower than the model prediction, it is not
a concern, since its abundance does not change much due to the
main component of the $s$-process \citep{pignatari2010,bisterzo2011,roederer2012c}.

For the first (Sr-Zr) and second (Ba-Nd) $s$-process peaks, all the abundances
are within $\sim$0.3~dex higher than the model. Other elements that are low
compared to the model are: Cd, Eu, Pt, and Au.  In the Solar System, 57\% of Cd
comes from the $s$-process \citep{arlandini1999}; this element is not expected
to be depleted.  This is also supported empirically by the sub-solar
\abund{Cd}{Fe} ratios found in low-metallicity stars with $r$-process material
\citep{roederer2014c}, which implies that another (secondary) process must
occur at later times or in higher-metallicity environments to bring the [Cd/Fe]
ratio up to the solar value.

For Eu, Au, and Pt, only  5-8\% of their abundances are produced by the
$s$-process.  Therefore, it is unexpected that the (small) amount of these
elements produced by the model with the $s$-process are higher than the
observed values.  This difference could be accounted for if we consider a
sub-solar abundance of Eu (and accordingly Au and Pt) at low metallicity.
However, it is worth mentioning that the previous Eu determination by
\citet{aoki2002} -- \eps{Eu}=$-$1.53 -- is about 0.5~dex higher than the one
from this work, and is in agreement with the models.  Further observations are
needed to reliably determine the abundances of these elements.  Nevertheless,
despite the discrepancies mentioned above, the overall model fits most of the
observed abundances within $\pm$0.5~dex.

We can compare our best-fitting model of HD 196944 with the previous results
obtained by \citet{abate2015b}. In this study, the same grid of
models was used as in the present work, but no derived orbital period
was available at the time to constrain the models. The orbital parameters found
 with model set B are $M_{1,i}=1.2 M_{\odot}$,
$M_{2,i}=0.79 M_{\odot}$, $P_{i}=6.9\times10^4$ days and $P_{f}=4.6\times10^4$
days. As a consequence of the period constraint, our best model has a lower
primary mass, a higher secondary mass, and shorter initial and final orbital
periods, namely $M_{1,i}=0.9 M_{\odot}$, $M_{2,i}=0.86 M_{\odot}$,
$P_{i}=1.64\times10^3$ days and $P_{f}=1.54\times10^3$ days. If we adopt a
higher primary mass of $M_{1,i}=1.2 M_{\odot}$, no model in our grid reproduces
the observed orbital period. Binary systems with initial periods shorter than
about $3300$ days experience a common-envelope phase as the primary star
overfills its Roche lobe during the AGB phase, and consequently the final
orbital periods are of the order of a few hundred days. Initially wider binary
systems do not lose enough angular momentum, thus their orbital periods
are much longer than the observed ($P_f>3000$ days). In contrast, if we ignore
the period constraint, our best-fitting model has essentially the same initial
parameters as \citet{abate2015b}, $M_{1,i}=1.2 M_{\odot}$, $M_{2,i}=0.79
M_{\odot}$, $P_{i}=9.8\times10^4$ days and $P_{f}=6.4\times10^4$ days. However,
the same regime of neutron-capture process operates in low-mass AGB stars
(M$_{AGB}\le1.5M_{\odot}$; e.g., Lugaro et al. 2012), and consequently the
$s$-element distributions do not vary signficantly between our models at
$M_{1,i}=1.2 M_{\odot}$ and $M_{1,i}=0.9 M_{\odot}$.

\subsubsection{HD~201626}

\hdb{} is a well-known CH star that has been extensively studied at optical
wavelengths -- SIMBAD notes 125 references.  Originally discovered to be a CH
star by \citet{northcott53}, the first spectroscopic chemical composition
analysis \citep{wg1963} identified \hdb{} as being rich in C, Ba, La, Ce, and
Nd.  The authors noted that the abundances of these species provided ``[...]
evidence for $s$-process element formation in old stars of low metal content.''
In this work we were able to add many new abundances for \hdb.  Particularly
important are the abundances of Na and Mg, since these place strong constraints
on the mass and the M$_{\rm PMZ}$ of the AGB donor star
\citep[e.g.,][]{abate2015a}.  Another important abundance determination is Eu,
which is a indicator of the operation of the $r$-process. With this abundance,
it is possible to calculate the ratio \abund{Ba}{Eu}=$+$0.78, which places
\hdb{} in the CEMP-$s$ sub-class, according to the the definition of
\citet{beers2005}.

The best-fit model calculated at metallicity $Z = 10^{-4}$ (\metal=$-$2.2;
upper panels of Figure~\ref{abateHD20}) shows good agreement with the
observations.  C, N, and O agree with the model within the observed
uncertainties.  Na, Mg, and Si are all within 0.05~dex from the model values.
However, some elements, such as Al, K, Sc, Ti, and V exhibit large residuals
when compared to the models.  For the elements in the first peak of the
$s$-process (Sr, Y, Zr, Mo), there is a large spread, but the abundances are
all within $\pm$0.5~dex from the model.  The other elements, in particular the
ones from the second peak of the $s$-process (Ba-Nd) are well-reproduced by the
model.  Similar to \hda, Pt is $\sim$0.4~dex lower than the model.  The Pb
abundance, which is also an important constraint of $s$-process
nucleosynthesis, is within $2 \sigma$ of the model values.

The results are significantly worse when the model
metallicity is $Z = 5\cdot10^{-4}$ (\metal=$-$1.5; lower panels of
Figure~\ref{abateHD20}), which would be, in principle, more appropriate for this
star. This is likely due to a limitation of the models, since these were optimized to
\metal=$-$2.2. The residuals for elements with $Z>38$ all shifted to higher values, 
meaning that the model is underproducing $s$-process elements. 

\citet{abate2015a} also studied the abundance pattern of \hdb, using 
chemical-abundance values from
the literature. The authors determined a higher primary mass ($M_{1,i} = 2.6
M_{\odot}$, with the same model set B) than our best-fit model ($M_{1,i} = 0.9
M_{\odot}$), because the abundances of Na and Mg were undetermined at the time.
Consequently, a relatively massive star with a large partial mixing zone
($M_{PMZ}=4\times10^{-3} M_{\odot}$) well-reproduced the large observed
enhancements of heavy-$s$ elements and lead. This model also predicts high Na
and Mg abundances ([Na/Fe]$_{mod} = +2.2$ and [Mg/Fe]$_{mod} = +1.5$), which are
ruled out by our new observations, [Na/Fe] = $+0.02$, and [Mg/Fe] = $+0.27$.

\section{Concluding Remarks}
\label{conc}

In this work we have presented abundance analyses of the NUV
HST/STIS spectra of two bright CEMP-$s$ stars, complemented by abundance
determinations from optical Keck/HIRES spectra. Abundances for some of the
elements measured in this work are only accessible from the NUV part of the
spectra, and are of great importance for testing theoretical predictions of
$s$-process nucleosynthesis. We also provided the first determination of the
orbital period for \hda. We compared our results with models of AGB
nucleosynthesis, as well as models for the evolution of binary systems
containing CEMP stars. Results are in good agreement with the models, and are
yet another confirmation of the formation scenario for the CEMP-$s$ stars.

To our knowledge, this is the first attempt to match theoretical predictions of
AGB evolution models and binary evolution with elemental abundances measured in the NUV.
This opens a new window of opportunity for further high-resolution spectroscopy
with HST/STIS for CEMP-$s$ stars, which will help further constrain 
not only specific characteristics of AGB stars, but also for the evolution of
the binary system in which these objects are found.  To carry out such an effort,
newly identified {\it bright} CEMP-$s$ stars are required.

\acknowledgments 

V.M.P., T.C.B., and H.S. acknowledge partial support for this work from PHY
08-22648; Physics Frontier Center/{}Joint Institute for Nuclear Astrophysics
(JINA) and PHY 14-30152; Physics Frontier Center/JINA Center for the Evolution
of the Elements (JINA-CEE), awarded by the US National Science Foundation.  T.
H. was supported by Sonderforschungsbereich SFB 881 "The Milky Way System"
(subproject A4) of the German Research Foundation (DFG) A.F. is supported by
NSF CAREER grant AST-1255160.  W.A. was supported by the JSPS Grants-in-Aid for
Scientific Research (23224004).  Generous support for Program GO-12554 has been
provided by NASA through a grant from the Space Telescope Science Institute,
which is operated by the Association of Universities for Research in Astronomy,
Inc., under NASA contract NAS 5-26555. This research has made use of the Keck
Observatory Archive, which is operated by the W.\ M.\ Keck Observatory and the
NASA Exoplanet Science Institute, under contract with the National Aeronautics
and Space Administration; the SIMBAD database, operated at CDS, Strasbourg,
France; and the NASA/IPAC Infrared Science Archive, which is operated by the
Jet Propulsion Laboratory, California Institue of Technology, under contract
with the National Aeronautics and Space Administration.
The authors also wish to recognize and acknowledge the very significant
cultural role and reverence that the summit of Mauna Kea has always had within
the indigenous Hawaiian community. We are most fortunate to have the
opportunity to conduct observations from this mountain.

{\it Facilities:} \facility{HST (STIS)}, \facility{Keck:II (HIRES)}

\bibliographystyle{apj}

\begin{thebibliography}{}
\expandafter\ifx\csname natexlab\endcsname\relax\def\natexlab#1{#1}\fi

\bibitem[{{Abate} {et~al.}(2013){Abate}, {Pols}, {Izzard}, {Mohamed}, \& {de
  Mink}}]{abate2013}
{Abate}, C., {Pols}, O.~R., {Izzard}, R.~G., {Mohamed}, S.~S., \& {de Mink},
  S.~E. 2013, \aap, 552, A26

\bibitem[Abate et al.(2015{\natexlab{a}})]{abate2015a} Abate, C., Pols, O.~R., Karakas, 
A.~I., \& Izzard, R.~G.\ 2015, \aap, 576, A118 

\bibitem[Abate et al.(2015{\natexlab{b}})]{abate2015b} Abate, C., Pols, O.~R., 
Izzard, R.~G., \& Karakas, A.~I.\ 2015, arXiv:1507.04662 

\bibitem[Abate et al.(2015{\natexlab{c}})]{abate2015c} Abate, C., Pols, O.~R., 
Stancliffe, R.~J., et al.\ 2015, arXiv:1507.04969 

\bibitem[{{Andersson} {et~al.}(2015){Andersson}, {Grumer}, {Ryde},
  {Blackwell-Whitehead}, {Hutton}, {Zou}, {J{\"o}nsson}, \&
  {Brage}}]{andersson2015}
{Andersson}, M., {Grumer}, J., {Ryde}, N., {et~al.} 2015, \apjs, 216, 2

\bibitem[{{Anderton}(2015)}]{THIMBLES}
{Anderton}, T.~R. 2015, PhD thesis, in prep.

\bibitem[Aoki et al.(2006)]{aoki2006} Aoki, W., Frebel, A., 
Christlieb, N., et al.\ 2006, \apj, 639, 897 

\bibitem[{{Aoki} {et~al.}(2002){Aoki}, {Ryan}, {Norris}, {Beers}, {Ando}, \&
  {Tsangarides}}]{aoki2002}
{Aoki}, W., {Ryan}, S.~G., {Norris}, J.~E., {et~al.} 2002, \apj, 580, 1149

\bibitem[{{Aoki} \& {Tsuji}(1997)}]{aoki1997}
{Aoki}, W., \& {Tsuji}, T. 1997, \aap, 317, 845

\bibitem[{{Arlandini} {et~al.}(1999){Arlandini}, {K{\"a}ppeler}, {Wisshak},
  {Gallino}, {Lugaro}, {Busso}, \& {Straniero}}]{arlandini1999}
{Arlandini}, C., {K{\"a}ppeler}, F., {Wisshak}, K., {et~al.} 1999, \apj, 525,
  886

\bibitem[{{Asplund}(2005)}]{asplund2005}
{Asplund}, M. 2005, \araa, 43, 481

\bibitem[{{Asplund} {et~al.}(2009){Asplund}, {Grevesse}, {Sauval}, \&
  {Scott}}]{asplund2009}
{Asplund}, M., {Grevesse}, N., {Sauval}, A.~J., \& {Scott}, P. 2009, \araa, 47,
  481

\bibitem[{{Barnbaum} {et~al.}(1996){Barnbaum}, {Stone}, \& {Keenan}}]{bsk1996}
{Barnbaum}, C., {Stone}, R.~P.~S., \& {Keenan}, P.~C. 1996, \apjs, 105, 419

\bibitem[{{Beers} \& {Christlieb}(2005)}]{beers2005}
{Beers}, T.~C., \& {Christlieb}, N. 2005, \araa, 43, 531

\bibitem[{{Bergeson} \& {Lawler}(1993)}]{bergeson1993}
{Bergeson}, S.~D., \& {Lawler}, J.~E. 1993, \apj, 408, 382

\bibitem[{{Bidelman}(1981)}]{bidelman1981}
{Bidelman}, W.~P. 1981, \aj, 86, 553

\bibitem[{{Bi{\'e}mont} {et~al.}(2000){Bi{\'e}mont}, {Garnir}, {Palmeri}, {Li},
  \& {Svanberg}}]{biemont2000}
{Bi{\'e}mont}, E., {Garnir}, H.~P., {Palmeri}, P., {Li}, Z.~S., \& {Svanberg},
  S. 2000, \mnras, 312, 116

\bibitem[{{Bisterzo} {et~al.}(2010){Bisterzo}, {Gallino}, {Straniero},
  {Cristallo}, \& {K{\"a}ppeler}}]{bisterzo2010}
{Bisterzo}, S., {Gallino}, R., {Straniero}, O., {Cristallo}, S., \&
  {K{\"a}ppeler}, F. 2010, \mnras, 404, 1529

\bibitem[{{Bisterzo} {et~al.}(2011){Bisterzo}, {Gallino}, {Straniero},
  {Cristallo}, \& {K{\"a}ppeler}}]{bisterzo2011}
---. 2011, \mnras, 418, 284

\bibitem[{{Bisterzo} {et~al.}(2012){Bisterzo}, {Gallino}, {Straniero},
  {Cristallo}, \& {K{\"a}ppeler}}]{bisterzo2012}
---. 2012, \mnras, 422, 849

\bibitem[{{Blackwell} {et~al.}(1980){Blackwell}, {Petford}, \&
  {Shallis}}]{bps1980}
{Blackwell}, D.~E., {Petford}, A.~D., \& {Shallis}, M.~J. 1980, \aap, 82, 249

\bibitem[{{Blackwell-Whitehead} \& {Bergemann}(2007)}]{bwb2007}
{Blackwell-Whitehead}, R., \& {Bergemann}, M. 2007, \aap, 472, L43

\bibitem[{{Buchhave} {et~al.}(2010){Buchhave}, {Bakos}, {Hartman}, {Torres},
  {Kov{\'a}cs}, {Latham}, {Noyes}, {Esquerdo}, {Everett}, {Howard}, {Marcy},
  {Fischer}, {Johnson}, {Andersen}, {F{\H u}r{\'e}sz}, {Perumpilly},
  {Sasselov}, {Stefanik}, {B{\'e}ky}, {L{\'a}z{\'a}r}, {Papp}, \&
  {S{\'a}ri}}]{buchhave2010}
{Buchhave}, L.~A., {Bakos}, G.~{\'A}., {Hartman}, J.~D., {et~al.} 2010, \apj,
  720, 1118

\bibitem[{{Calamida} {et~al.}(2007){Calamida}, {Bono}, {Stetson}, {Freyhammer},
  {Cassisi}, {Grundahl}, {Pietrinferni}, {Hilker}, {Primas}, {Richtler},
  {Romaniello}, {Buonanno}, {Caputo}, {Castellani}, {Corsi}, {Ferraro},
  {Iannicola}, \& {Pulone}}]{calamida+2007}
{Calamida}, A., {Bono}, G., {Stetson}, P.~B., {et~al.} 2007, \apj, 670, 400

\bibitem[{{Casagrande} {et~al.}(2011){Casagrande}, {Schonrich}, {Asplund},
  {Cassisi}, {Ramirez}, {Melendez}, {Bensby}, \& {Feltzing}}]{casagrande+2011}
{Casagrande}, L., {Schonrich}, R., {Asplund}, M., {et~al.} 2011, \aap, 530,
  138+21

\bibitem[{{Castelli} \& {Kurucz}(2004)}]{castelli2004}
{Castelli}, F., \& {Kurucz}, R.~L. 2004, ArXiv Astrophysics e-prints,
  arXiv:astro-ph/0405087

\bibitem[{{Collet} {et~al.}(2007){Collet}, {Asplund}, \&
  {Trampedach}}]{collet2007}
{Collet}, R., {Asplund}, M., \& {Trampedach}, R. 2007, \aap, 469, 687

\bibitem[{{Cowan} {et~al.}(2005){Cowan}, {Sneden}, {Beers}, {Lawler},
  {Simmerer}, {Truran}, {Primas}, {Collier}, \& {Burles}}]{cowan2005}
{Cowan}, J.~J., {Sneden}, C., {Beers}, T.~C., {et~al.} 2005, \apj, 627, 238

\bibitem[{{Cristallo} {et~al.}(2011){Cristallo}, {Piersanti}, {Straniero},
  {Gallino}, {Dom{\'{\i}}nguez}, {Abia}, {Di Rico}, {Quintini}, \&
  {Bisterzo}}]{cristallo2011}
{Cristallo}, S., {Piersanti}, L., {Straniero}, O., {et~al.} 2011, \apjs, 197,
  17

\bibitem[{{Den Hartog} {et~al.}(2005){Den Hartog}, {Herd}, {Lawler}, {Sneden},
  {Cowan}, \& {Beers}}]{denhartog2005}
{Den Hartog}, E.~A., {Herd}, M.~T., {Lawler}, J.~E., {et~al.} 2005, \apj, 619,
  639

\bibitem[{{Den Hartog} {et~al.}(2011){Den Hartog}, {Lawler}, {Sobeck},
  {Sneden}, \& {Cowan}}]{denhartog2011}
{Den Hartog}, E.~A., {Lawler}, J.~E., {Sobeck}, J.~S., {Sneden}, C., \&
  {Cowan}, J.~J. 2011, \apjs, 194, 35

\bibitem[{{Fabbian} {et~al.}(2009){Fabbian}, {Asplund}, {Barklem}, {Carlsson},
  \& {Kiselman}}]{fabian+2009}
{Fabbian}, D., {Asplund}, M., {Barklem}, P.~S., {Carlsson}, M., \& {Kiselman},
  D. 2009, \aap, 500, 1221

\bibitem[{{Fedchak} \& {Lawler}(1999)}]{fedchak1999}
{Fedchak}, J.~A., \& {Lawler}, J.~E. 1999, \apj, 523, 734

\bibitem[{{Fitzpatrick} \& {Sneden}(1987)}]{SPECTRE}
{Fitzpatrick}, M.~J., \& {Sneden}, C. 1987, in Bulletin of the American
  Astronomical Society, Vol.~19, Bulletin of the American Astronomical Society,
  1129

\bibitem[{{Fivet} {et~al.}(2006){Fivet}, {Quinet}, {Bi{\'e}mont}, \&
  {Xu}}]{fivet2006}
{Fivet}, V., {Quinet}, P., {Bi{\'e}mont}, {\'E}., \& {Xu}, H.~L. 2006, Journal
  of Physics B Atomic Molecular Physics, 39, 3587

\bibitem[{{Frebel} \& {Norris}(2015)}]{frebel2015}
{Frebel}, A., \& {Norris}, J.~E. 2015, ArXiv e-prints, arXiv:1501.06921

\bibitem[{{Fuhr} \& {Wiese}(2009)}]{fuhr2009}
{Fuhr}, J.~R., \& {Wiese}, W.~L. 2009, published in the CRC Handbook of
  Chemistry and Physics

\bibitem[{{Garcia Perez} {et~al.}(2006){Garcia Perez}, {Asplund}, {Primas},
  {Nissen}, \& {Gustafsson}}]{garciaperez2006}
{Garcia Perez}, A.~E., {Asplund}, M., {Primas}, F., {Nissen}, P.~E., \&
  {Gustafsson}, G. 2006, \aap, 451, 621

\bibitem[{{Hauck} \& {Mermilliod}(1997)}]{hauck1997}
{Hauck}, B., \& {Mermilliod}, M. 1997, VizieR Online Data Catalog, 2215, 0

\bibitem[{{Ivans} {et~al.}(2006){Ivans}, {Simmerer}, {Sneden}, {Lawler},
  {Cowan}, {Gallino}, \& {Bisterzo}}]{ivans2006}
{Ivans}, I.~I., {Simmerer}, J., {Sneden}, C., {et~al.} 2006, \apj, 645, 613

\bibitem[{{Ivans} {et~al.}(2005){Ivans}, {Sneden}, {Gallino}, {Cowan}, \&
  {Preston}}]{ivans+2005}
{Ivans}, I.~I., {Sneden}, C., {Gallino}, R., {Cowan}, J.~C., \& {Preston},
  G.~W. 2005, \apj, 627, L145

\bibitem[{{Ivans} {et~al.}(2003){Ivans}, {Sneden}, {James}, {Preston},
  {Fulbright}, {H{\"o}flich}, {Carney}, \& {Wheeler}}]{ivans2003}
{Ivans}, I.~I., {Sneden}, C., {James}, C.~R., {et~al.} 2003, \apj, 592, 906

\bibitem[{{Karakas}(2010)}]{karakas2010}
{Karakas}, A.~I. 2010, \mnras, 403, 1413

\bibitem[{{Karakas} {et~al.}(2009){Karakas}, {van Raai}, {Lugaro}, {Sterling},
  \& {Dinerstein}}]{karakas2009}
{Karakas}, A.~I., {van Raai}, M.~A., {Lugaro}, M., {Sterling}, N.~C., \&
  {Dinerstein}, H.~L. 2009, \apj, 690, 1130

\bibitem[{{Karinkuzhi} \& {Goswami}(2014)}]{karinkuzhi2014}
{Karinkuzhi}, D., \& {Goswami}, A. 2014, \mnras, 440, 1095

\bibitem[{{Kimble} {et~al.}(1998){Kimble}, {Woodgate}, {Bowers}, {Kraemer},
  {Kaiser}, {Gull}, {Heap}, {Danks}, {Boggess}, {Green}, {Hutchings},
  {Jenkins}, {Joseph}, {Linsky}, {Maran}, {Moos}, {Roesler}, {Timothy},
  {Weistrop}, {Grady}, {Loiacono}, {Brown}, {Brumfield}, {Content}, {Feinberg},
  {Isaacs}, {Krebs}, {Krueger}, {Melcher}, {Rebar}, {Vitagliano}, {Yagelowich},
  {Meyer}, {Hood}, {Argabright}, {Becker}, {Bottema}, {Breyer}, {Bybee},
  {Christon}, {Delamere}, {Dorn}, {Downey}, {Driggers}, {Ebbets}, {Gallegos},
  {Garner}, {Hetlinger}, {Lettieri}, {Ludtke}, {Michika}, {Nyquist}, {Rose},
  {Stocker}, {Sullivan}, {van Houten}, {Woodruff}, {Baum}, {Hartig}, {Balzano},
  {Biagetti}, {Blades}, {Bohlin}, {Clampin}, {Doxsey}, {Ferguson},
  {Goudfrooij}, {Hulbert}, {Kutina}, {McGrath}, {Lindler}, {Beck}, {Feggans},
  {Plait}, {Sandoval}, {Hill}, {Collins}, {Cornett}, {Fowler}, {Hill},
  {Landsman}, {Malumuth}, {Standley}, {Blouke}, {Grusczak}, {Reed}, {Robinson},
  {Valenti}, \& {Wolfe}}]{kimble1998}
{Kimble}, R.~A., {Woodgate}, B.~E., {Bowers}, C.~W., {et~al.} 1998, \apjl, 492,
  L83

\bibitem[{{Kobayashi} {et~al.}(2011){Kobayashi}, {Karakas}, \&
  {Umeda}}]{kobayashi2011c}
{Kobayashi}, C., {Karakas}, A.~I., \& {Umeda}, H. 2011, \mnras, 414, 3231

\bibitem[{{Kraft} \& {Ivans}(2003)}]{KI2003}
{Kraft}, R.~P., \& {Ivans}, I.~I. 2003, \pasp, 115, 143

\bibitem[{{Kraft} \& {Ivans}(2004)}]{KI2004}
---. 2004, Origin and Evolution of the Elements, 33

\bibitem[{Kramida {et~al.}(2013)Kramida, {Yu.~Ralchenko}, Reader, \& {and NIST
  ASD Team}}]{nist}
Kramida, A., {Yu.~Ralchenko}, Reader, J., \& {and NIST ASD Team}. 2013, {NIST
  Atomic Spectra Database (ver. 5.1), [Online]. Available:
  {\tt{http://physics.nist.gov/asd}} [2014, April 7]. National Institute of
  Standards and Technology, Gaithersburg, MD.}

\bibitem[{{Kupka} {et~al.}(1999){Kupka}, {Piskunov}, {Ryabchikova}, {Stempels},
  \& {Weiss}}]{vald}
{Kupka}, F., {Piskunov}, N., {Ryabchikova}, T.~A., {Stempels}, H.~C., \&
  {Weiss}, W.~W. 1999, \aaps, 138, 119

\bibitem[Kurucz(1993)]{kurucz} Kurucz, R.~L.\ 1993, Kurucz 
CD-ROM, Cambridge, MA: Smithsonian Astrophysical Observatory, |c1993, December 4, 1993

\bibitem[{{Kurucz} \& {Bell}(1995)}]{kurucz1995}
{Kurucz}, R., \& {Bell}, B. 1995, Atomic Line Data (R.L.~Kurucz and B.~Bell)
  Kurucz CD-ROM No.~23.~Cambridge, Mass.: Smithsonian Astrophysical
  Observatory, 1995., 23

\bibitem[{{Lawler} {et~al.}(2007){Lawler}, {den Hartog}, {Labby}, {Sneden},
  {Cowan}, \& {Ivans}}]{lawler2007}
{Lawler}, J.~E., {den Hartog}, E.~A., {Labby}, Z.~E., {et~al.} 2007, \apjs,
  169, 120

\bibitem[{{Lawler} {et~al.}(2014){Lawler}, {Wood}, {Den Hartog}, {Feigenson},
  {Sneden}, \& {Cowan}}]{lawler2014}
{Lawler}, J.~E., {Wood}, M.~P., {Den Hartog}, E.~A., {et~al.} 2014, \apjs, 215,
  20

\bibitem[{{Ljung} {et~al.}(2006){Ljung}, {Nilsson}, {Asplund}, \&
  {Johansson}}]{ljung2006}
{Ljung}, G., {Nilsson}, H., {Asplund}, M., \& {Johansson}, S. 2006, \aap, 456,
  1181

\bibitem[{{Lucatello} {et~al.}(2005){Lucatello}, {Tsangarides}, {Beers},
  {Carretta}, {Gratton}, \& {Ryan}}]{lucatello2005}
{Lucatello}, S., {Tsangarides}, S., {Beers}, T.~C., {et~al.} 2005, \apj, 625,
  825

\bibitem[{{Lugaro} {et~al.}(2012){Lugaro}, {Karakas}, {Stancliffe}, \&
  {Rijs}}]{lugaro2012}
{Lugaro}, M., {Karakas}, A.~I., {Stancliffe}, R.~J., \& {Rijs}, C. 2012, \apj,
  747, 2

\bibitem[{{Malcheva} {et~al.}(2006){Malcheva}, {Blagoev}, {Mayo}, {Ortiz},
  {Xu}, {Svanberg}, {Quinet}, \& {Bi{\'e}mont}}]{malcheva2006}
{Malcheva}, G., {Blagoev}, K., {Mayo}, R., {et~al.} 2006, \mnras, 367, 754

\bibitem[{{Mashonkina} {et~al.}(2012){Mashonkina}, {Ryabtsev}, \&
  {Frebel}}]{mashonkina2012}
{Mashonkina}, L., {Ryabtsev}, A., \& {Frebel}, A. 2012, \aap, 540, A98

\bibitem[McClure \& Woodsworth(1990)]{mcclure1990} McClure, R.~D., \& 
Woodsworth, A.~W.\ 1990, \apj, 352, 709

\bibitem[{{McDonald} {et~al.}(2012){McDonald}, {Zijlstra}, \&
  {Boyer}}]{mzb2012}
{McDonald}, I., {Zijlstra}, A.~A., \& {Boyer}, M.~L. 2012, \mnras, 427, 343

\bibitem[{{Mel{\'e}ndez} \& {Barbuy}(2009)}]{melendez2009b}
{Mel{\'e}ndez}, J., \& {Barbuy}, B. 2009, \aap, 497, 611

\bibitem[{{Morton}(2000)}]{morton2000}
{Morton}, D.~C. 2000, \apjs, 130, 403

\bibitem[{{Nilsson} \& {Ivarsson}(2008)}]{nilsson2008}
{Nilsson}, H., \& {Ivarsson}, S. 2008, \aap, 492, 609

\bibitem[{{Nordstr{\"o}m} {et~al.}(2004){Nordstr{\"o}m}, {Mayor}, {Andersen},
  {Holmberg}, {Pont}, {J{\o}rgensen}, {Olsen}, {Udry}, \&
  {Mowlavi}}]{nordstrom2004}
{Nordstr{\"o}m}, B., {Mayor}, M., {Andersen}, J., {et~al.} 2004, \aap, 418, 989

\bibitem[{{Northcott}(1953)}]{northcott53}
{Northcott}, R.~J. 1953, \jrasc, 47, 65

\bibitem[{{P{\'a}l}(2009)}]{pal2009}
{P{\'a}l}, A. 2009, \mnras, 396, 1737

\bibitem[{{Pignatari} {et~al.}(2010){Pignatari}, {Gallino}, {Heil}, {Wiescher},
  {K{\"a}ppeler}, {Herwig}, \& {Bisterzo}}]{pignatari2010}
{Pignatari}, M., {Gallino}, R., {Heil}, M., {et~al.} 2010, \apj, 710, 1557

\bibitem[{{Placco} {et~al.}(2013){Placco}, {Frebel}, {Beers}, {Karakas},
  {Kennedy}, {Rossi}, {Christlieb}, \& {Stancliffe}}]{placco2013}
{Placco}, V.~M., {Frebel}, A., {Beers}, T.~C., {et~al.} 2013, \apj, 770, 104

\bibitem[{{Placco} {et~al.}(2014{\natexlab{a}}){Placco}, {Frebel}, {Beers}, \&
  {Stancliffe}}]{placco2014c}
{Placco}, V.~M., {Frebel}, A., {Beers}, T.~C., \& {Stancliffe}, R.~J.
  2014{\natexlab{a}}, \apj, 797, 21

\bibitem[{{Placco} {et~al.}(2014{\natexlab{b}}){Placco}, {Beers}, {Roederer},
  {Cowan}, {Frebel}, {Filler}, {Ivans}, {Lawler}, {Schatz}, {Sneden}, {Sobeck},
  {Aoki}, \& {Smith}}]{placco2014b}
{Placco}, V.~M., {Beers}, T.~C., {Roederer}, I.~U., {et~al.}
  2014{\natexlab{b}}, \apj, 790, 34

\bibitem[{{Quinet} {et~al.}(2006){Quinet}, {Palmeri}, {Bi{\'e}mont},
  {Jorissen}, {van Eck}, {Svanberg}, {Xu}, \& {Plez}}]{quinet2006}
{Quinet}, P., {Palmeri}, P., {Bi{\'e}mont}, {\'E}., {et~al.} 2006, \aap, 448,
  1207

\bibitem[{{Ramirez} {et~al.}(2013){Ramirez}, {Allende Prieto}, \&
  {Lambert}}]{ramirez+2013}
{Ramirez}, I., {Allende Prieto}, C., \& {Lambert}, D.~L. 2013, \apj, 764, 78

\bibitem[{{Ramirez} \& {Melendez}(2005)}]{rm2005}
{Ramirez}, I., \& {Melendez}, J. 2005, \apj, 743, 465

\bibitem[{{Roederer}(2012)}]{roederer2012c}
{Roederer}, I.~U. 2012, \apj, 756, 36

\bibitem[{{Roederer} {et~al.}(2010){Roederer}, {Cowan}, {Karakas}, {Kratz},
  {Lugaro}, {Simmerer}, {Farouqi}, \& {Sneden}}]{roederer2010b}
{Roederer}, I.~U., {Cowan}, J.~J., {Karakas}, A.~I., {et~al.} 2010, \apj, 724,
  975

\bibitem[{{Roederer} {et~al.}(2009){Roederer}, {Kratz}, {Frebel}, {Christlieb},
  {Pfeiffer}, {Cowan}, \& {Sneden}}]{roederer2009}
{Roederer}, I.~U., {Kratz}, K., {Frebel}, A., {et~al.} 2009, \apj, 698, 1963

\bibitem[{{Roederer} {et~al.}(2008){Roederer}, {Lawler}, {Sneden}, {Cowan},
  {Sobeck}, \& {Pilachowski}}]{roederer2008}
{Roederer}, I.~U., {Lawler}, J.~E., {Sneden}, C., {et~al.} 2008, \apj, 675, 723

\bibitem[{{Roederer} {et~al.}(2014{\natexlab{a}}){Roederer}, {Preston},
  {Thompson}, {Shectman}, {Sneden}, {Burley}, \& {Kelson}}]{roederer2014}
{Roederer}, I.~U., {Preston}, G.~W., {Thompson}, I.~B., {et~al.}
  2014{\natexlab{a}}, \aj, 147, 136

\bibitem[{{Roederer} {et~al.}(2012){Roederer}, {Lawler}, {Sobeck}, {Beers},
  {Cowan}, {Frebel}, {Ivans}, {Schatz}, {Sneden}, \&
  {Thompson}}]{roederer2012d}
{Roederer}, I.~U., {Lawler}, J.~E., {Sobeck}, J.~S., {et~al.} 2012, \apjs, 203,
  27

\bibitem[{{Roederer} {et~al.}(2014{\natexlab{b}}){Roederer}, {Schatz},
  {Lawler}, {Beers}, {Cowan}, {Frebel}, {Ivans}, {Sneden}, \&
  {Sobeck}}]{roederer2014c}
{Roederer}, I.~U., {Schatz}, H., {Lawler}, J.~E., {et~al.} 2014{\natexlab{b}},
  \apj, 791, 32

\bibitem[{{Ruffoni} {et~al.}(2014){Ruffoni}, {Den Hartog}, {Lawler}0, {Brewer},
  {Lind}, {Nave}, \& {Pickering}}]{ruffoni+14}
{Ruffoni}, M.~P., {Den Hartog}, E.~A., {Lawler}0, J.~E., {et~al.} 2014, \mnras,
  44, 3127

\bibitem[{{Schlafly} \& {Finkbeiner}(2011)}]{sf2011}
{Schlafly}, E.~F., \& {Finkbeiner}, D.~P. 2011, \apj, 737, 103

\bibitem[{{Sikstr{\"o}m} {et~al.}(2001){Sikstr{\"o}m}, {Pihlemark}, {Nilsson},
  {Litz{\'e}n}, {Johansson}, {Li}, \& {Lundberg}}]{sikstrom2001}
{Sikstr{\"o}m}, C.~M., {Pihlemark}, H., {Nilsson}, H., {et~al.} 2001, Journal
  of Physics B Atomic Molecular Physics, 34, 477

\bibitem[Simmerer et al.(2013)]{simmerer2013} Simmerer, J., Ivans, 
I.~I., Filler, D., et al.\ 2013, \apjl, 764, L7

\bibitem[Siqueira Mello et al.(2013)]{siqueira2013} Siqueira Mello, 
C., Spite, M., Barbuy, B., et al.\ 2013, \aap, 550, A122 


\bibitem[{{Sneden} {et~al.}(2014){Sneden}, {Lucatello}, {Ram}, {Brooke}, \&
  {Bernath}}]{sneden2014}
{Sneden}, C., {Lucatello}, S., {Ram}, R.~S., {Brooke}, J.~S.~A., \& {Bernath},
  P. 2014, \apjs, 214, 26

\bibitem[{{Sneden} {et~al.}(2003){Sneden}, {Cowan}, {Lawler}, {Ivans},
  {Burles}, {Beers}, {Primas}, {Hill}, {Truran}, {Fuller}, {Pfeiffer}, \&
  {Kratz}}]{sneden2003}
{Sneden}, C., {Cowan}, J.~J., {Lawler}, J.~E., {et~al.} 2003, \apj, 591, 936

\bibitem[{{Sneden}(1973)}]{sneden1973}
{Sneden}, C.~A. 1973, PhD thesis, The University of Texas at Austin.

\bibitem[{{Sobeck} {et~al.}(2011){Sobeck}, {Kraft}, {Sneden}, {Preston},
  {Cowan}, {Smith}, {Thompson}, {Shectman}, \& {Burley}}]{sobeck2011}
{Sobeck}, J.~S., {Kraft}, R.~P., {Sneden}, C., {et~al.} 2011, \aj, 141, 175

\bibitem[{{Takeda}(2003)}]{takeda2003}
{Takeda}, Y. 2003, \aap, 402, 343

\bibitem[{{Van Eck} {et~al.}(2003){Van Eck}, {Goriely}, {Jorissen}, \&
  {Plez}}]{vaneck2003}
{Van Eck}, S., {Goriely}, S., {Jorissen}, A., \& {Plez}, B. 2003, \aap, 404,
  291

\bibitem[{{Vanture}(1992)}]{vanture1992}
{Vanture}, A.~D. 1992, \aj, 103, 2035

\bibitem[{{Vogt} {et~al.}(1994){Vogt}, {Allen}, {Bigelow}, {Bresee}, {Brown},
  {Cantrall}, {Conrad}, {Couture}, {Delaney}, {Epps}, {Hilyard}, {Hilyard},
  {Horn}, {Jern}, {Kanto}, {Keane}, {Kibrick}, {Lewis}, {Osborne},
  {Pardeilhan}, {Pfister}, {Ricketts}, {Robinson}, {Stover}, {Tucker}, {Ward},
  \& {Wei}}]{vogt1994}
{Vogt}, S.~S., {Allen}, S.~L., {Bigelow}, B.~C., {et~al.} 1994, in Society of
  Photo-Optical Instrumentation Engineers (SPIE) Conference Series, Vol. 2198,
  Instrumentation in Astronomy VIII, ed. D.~L. {Crawford} \& E.~R. {Craine},
  362

\bibitem[{{Wallerstein} \& {Dominy}(1988)}]{wd1988}
{Wallerstein}, G., \& {Dominy}, J.~F. 1988, \apj, 330, 937

\bibitem[{{Wallerstein} \& {Greenstein}(1963)}]{wg1963}
{Wallerstein}, G., \& {Greenstein}, J.~L. 1963, \aj, 68, 546

\bibitem[{{Wallerstein} \& {Greenstein}(1964)}]{wg1964}
---. 1964, \apj, 139, 1163

\bibitem[{{Waters} \& {Hollek}(2013)}]{robospect}
{Waters}, C.~Z., \& {Hollek}, J.~K. 2013, \pasp, 125, 1164

\bibitem[{{Wenger} {et~al.}(2000){Wenger}, {Ochsenbein}, {Egret}, {Dubois},
  {Bonnarel}, {Borde}, {Genova}, {Jasniewicz}, {Lalo{\"e}}, {Lesteven}, \&
  {Monier}}]{SIMBAD}
{Wenger}, M., {Ochsenbein}, F., {Egret}, D., {et~al.} 2000, \aaps, 143, 9

\bibitem[{{Wood} {et~al.}(2014{\natexlab{a}}){Wood}, {Lawler}, {Den Hartog},
  {Sneden}, \& {Cowan}}]{wood2014a}
{Wood}, M.~P., {Lawler}, J.~E., {Den Hartog}, E.~A., {Sneden}, C., \& {Cowan},
  J.~J. 2014{\natexlab{a}}, \apjs, 214, 18

\bibitem[{{Wood} {et~al.}(2013){Wood}, {Lawler}, {Sneden}, \&
  {Cowan}}]{wood2013}
{Wood}, M.~P., {Lawler}, J.~E., {Sneden}, C., \& {Cowan}, J.~J. 2013, \apjs,
  208, 27

\bibitem[{{Wood} {et~al.}(2014{\natexlab{b}}){Wood}, {Lawler}, {Sneden}, \&
  {Cowan}}]{wood2014b}
---. 2014{\natexlab{b}}, \apjs, 211, 20

\bibitem[{{Woodgate} {et~al.}(1998){Woodgate}, {Kimble}, {Bowers}, {Kraemer},
  {Kaiser}, {Danks}, {Grady}, {Loiacono}, {Brumfield}, {Feinberg}, {Gull},
  {Heap}, {Maran}, {Lindler}, {Hood}, {Meyer}, {Vanhouten}, {Argabright},
  {Franka}, {Bybee}, {Dorn}, {Bottema}, {Woodruff}, {Michika}, {Sullivan},
  {Hetlinger}, {Ludtke}, {Stocker}, {Delamere}, {Rose}, {Becker}, {Garner},
  {Timothy}, {Blouke}, {Joseph}, {Hartig}, {Green}, {Jenkins}, {Linsky},
  {Hutchings}, {Moos}, {Boggess}, {Roesler}, \& {Weistrop}}]{woodgate1998}
{Woodgate}, B.~E., {Kimble}, R.~A., {Bowers}, C.~W., {et~al.} 1998, \pasp, 110,
  1183

\bibitem[{{Zacs} {et~al.}(1998){Zacs}, {Nissen}, \& {Schuster}}]{zacs1998}
{Zacs}, L., {Nissen}, P.~E., \& {Schuster}, W.~J. 1998, \aap, 337, 216

\bibitem[{{Zinn} \& {West}(1984)}]{zw1984}
{Zinn}, R., \& {West}, M.~J. 1984, \apjs, 55, 45

\end{thebibliography}

\clearpage

\ltab

\begin{deluxetable}{lcclrl}
\tabletypesize{\scriptsize}
\tablewidth{0pc}
\tablecaption{Atmospheric Parameters\label{atmlit}}
\tablehead{
\colhead{\teff}&
\colhead{\logg}&
\colhead{\metal}&
\colhead{$\xi$}&
\colhead{Method}&
\colhead{Reference}\\
\colhead{(K)}&
\colhead{(cgs)}&
\colhead{}&
\colhead{(km~s$^{-1}$)}&
\colhead{}&
\colhead{}\\}
\startdata
\multicolumn{6}{c}{\protect\hda\ (literature)} \\
\hline
5170    & 1.60    & $-$2.41    & 1.55    &   a & This work            \\
5250    & 1.70    & $-$2.45    & 1.90    &   a & \citet{zacs1998}     \\
5250    & 1.80    & $-$2.25    & 1.70    &   a & \citet{aoki2002}     \\
\nodata & \nodata & $-$2.40    & \nodata &   a & \citet{vaneck2003}   \\
5170    & 1.80    & $-$2.46    & 1.70    &   a & \citet{roederer2008} \\
5310    & 1.75    & $-$2.39    & 1.65    &   a & \citet{roederer2014} \\
\hline
\multicolumn{6}{c}{\protect\hdb\ (literature)} \\
\enddata
\tablenotetext{*}{
Estimates made by adopting the $E(B-V)$ value based on the dust maps of
\citet{sf2011} at the IRSA Galactic Dust Reddening and Extinction
Service\footnote{ \href{http://irsa.ipac.caltech.edu/applications/DUST/}
{http://irsa.ipac.caltech.edu/applications/DUST/}}.  However, as noted by
\citet{sneden2014}, because the star is only some 300 pc away, one would expect
to see a smaller dereddening effect from a fraction of the total $E(B-V)$ value
(and thus, an even cooler photometric color-\teff).
}
\tablecomments{
	\\
	(a) Spectroscopic constraints \\
	(b) IRFM \citep{bps1980} \\
	(c) color-Teff \citep{rm2005} \\
	(d) color-Teff \citep{casagrande+2011} \\
	(e) SED modelling \\
	(f) $uvby$-metallicity \citep{calamida+2007}; empirical relation \\
	(g) $uvby$-metallicity \citep{calamida+2007}; theoretical relation \\
	(h) converted to \citet{KI2003,KI2004} metallicity scale }
\end{deluxetable}

\begin{deluxetable}{lccrrr}
\tablecolumns{6} 
\tablewidth{0pc} 
\tabletypesize{\scriptsize}
\tablecaption{Derived Optical Abundance Estimates\label{abtableOP}} 
\tablehead{ 
\colhead{Ion}& 
\colhead{$\lambda$}& 
\colhead{$\chi$} &  
\colhead{$\log\,gf$}& 
\colhead{$\log\epsilon$\,(X)}&
\colhead{$\log\epsilon$\,(X)}\\
\colhead{}& 
\colhead{({\AA})}& 
\colhead{(eV)} &  
\colhead{}& 
\colhead{\hda}& 
\colhead{\hdb}} 
\startdata 

\multicolumn{6}{c}{Equivalent width analysis} \\
\hline

\ion{C}{1}  & 4771.75 & 7.49 & $-$1.87 & \nodata &    8.56 \\
\ion{C}{1}  & 5380.34 & 7.68 & $-$1.62 & \nodata &    8.26 \\
\ion{C}{1}  & 6587.61 & 8.54 & $-$1.00 & \nodata &    8.11 \\
\ion{Na}{1} & 5682.63 & 2.10 & $-$0.70 & $<$4.10 & \nodata \\
\ion{Na}{1} & 5688.19 & 2.10 & $-$1.40 &    3.90 &    4.50 \\
\ion{Na}{1} & 6160.75 & 2.10 & $-$1.25 & \nodata &    5.01 \\
\ion{Mg}{1} & 4571.10 & 0.00 & $-$5.57 &    5.34 &    6.39 \\
\enddata
\end{deluxetable}

\begin{deluxetable}{lccrrrr}
\tablecolumns{7} 
\tablewidth{0pc} 
\tabletypesize{\scriptsize}
\tablecaption{NUV Abundance Estimates and Upper Limits\label{abtable}} 
\tablehead{ 
\colhead{Ion}& 
\colhead{$\lambda$}& 
\colhead{$\chi$} &  
\colhead{$\log\,gf$}& 
\colhead{$\log\epsilon$\,(X)}&
\colhead{$\log\epsilon$\,(X)}&
\colhead{Ref.}\\
\colhead{}& 
\colhead{({\AA})}& 
\colhead{(eV)} &  
\colhead{}& 
\colhead{\hda}& 
\colhead{\hdb}& 
\colhead{}} 
\startdata 

\multicolumn{7}{c}{Equivalent width analysis} \\
\hline
\ion{Ti}{1} & 2941.99 & 0.00  & $-$0.14 &     3.00    & \nodata    &  1 \\ 
\ion{Ti}{1} & 2956.13 & 0.05  & $+$0.12 &     2.95    & \nodata    &  1 \\ 
\ion{Ti}{2} & 2716.25 & 1.08  & $-$1.48 &     3.18    & \nodata    &  1 \\ 
\ion{Ti}{2} & 2812.02 & 1.16  & $-$2.32 &     2.94    & \nodata    &  1 \\ 
\ion{Ti}{2} & 2832.18 & 0.57  & $-$0.85 &     3.12    & \nodata    &  1 \\ 
\ion{Ti}{2} & 2884.10 & 1.13  & $-$0.23 &     2.67    & \nodata    &  1 \\ 
\ion{Ti}{2} & 2888.93 & 0.57  & $-$1.36 &     3.11    & \nodata    &  1 \\
\enddata
\tablerefs{
(1)  \citet{wood2013};
(2)  \citet{bergeson1993};
(3)  \citet{wood2014a};
(4)  \citet{fedchak1999};
(5)  \citet{denhartog2011};
(6)  \citet{kurucz1995};
(7)  \citet{fuhr2009};
(8)  \citet{ljung2006};
(9)  \citet{malcheva2006};
(10  \citet{nilsson2008};
(11) \citet{sikstrom2001};
(12) \citet{morton2000};
(13) \citet{roederer2010b};
(14) \citet{lawler2007};
(15) \citet{quinet2006};
(16) \citet{denhartog2005};
(17) \citet{fivet2006};
(18) \citet{biemont2000}, using hfs presented in Appendix C of
     \citet{roederer2012d}.
}

\end{deluxetable}

\clearpage

\begin{deluxetable}{lrrrrrrrrrr}
\tablewidth{0pc}  
\tabletypesize{\scriptsize}
\tablecaption{Final Optical LTE Abundance Estimates for HD~196944 and HD~201626\label{abfinal2}} 
\tablehead{ 
&& \multicolumn{4}{c}{\hda} &  & \multicolumn{4}{c}{\hdb} \\
 \cline{3-6} \cline{8-11}
Species & $\log\epsilon_{\odot}$\,(X) & $\log\epsilon$\,(X) 
                                      & $\mbox{[X/Fe]}$ & $\sigma$ & $N$ &
                                      & $\log\epsilon$\,(X) 
									  & $\mbox{[X/Fe]}$ & $\sigma$ & $N$ }
\startdata 
C (CH)      & 8.43 &    7.10 & $+$1.08\xx &    0.20 &       2  &&    8.40 & $+$1.48\yy &    0.20 &       2 \\
N (NH)      & 7.83 &    6.50 & $+$1.08    &    0.25 &       1  &&    6.70 & $+$0.38    &    0.30 &       1 \\
\ion{O}{1}  & 8.69 & \nodata & \nodata    & \nodata & \nodata  &&    7.91 & $+$0.73    &    0.20 &       1 \\
\ion{Na}{1} & 6.24 &    4.65 & $+$0.82    &    0.01 &       2  &&    4.75 & $+$0.02    &    0.10 &       2 \\
\ion{Mg}{1} & 7.60 &    5.55 & $+$0.36    &    0.07 &       6  &&    6.36 & $+$0.27    &    0.08 &       4 \\
\ion{Al}{1} & 6.45 & \nodata & \nodata    & \nodata & \nodata  &&    6.19 & $+$1.25    &    0.07 &       2 \\
\ion{Si}{1} & 7.51 & \nodata & \nodata    & \nodata & \nodata  &&    6.32 & $+$0.32    &    0.03 &      27 \\
\ion{K}{1}  & 5.03 & \nodata & \nodata    & \nodata & \nodata  &&    4.50 & $+$0.98    &    0.20 &       1 \\
\enddata
\tablenotetext{a}{\cfe$>+$1.31 using corrections of \citet{placco2014c}.}
\tablenotetext{b}{\cfe=$+$1.50 using corrections of \citet{placco2014c}.}
\tablenotetext{c}{$\mbox{[FeI/H]}$ and $\mbox{[FeII/H]}$ values}
\end{deluxetable}

\begin{deluxetable}{lrrrrrrrrrr}
\tablewidth{0pc}  
\tabletypesize{\scriptsize}
\tablecaption{Final NUV LTE Abundance Estimates for HD~196944 and HD~201626\label{abfinal}} 
\tablehead{ 
&& \multicolumn{4}{c}{\hda} &  & \multicolumn{4}{c}{\hdb} \\
 \cline{3-6} \cline{8-11}
Species & $\log\epsilon_{\odot}$\,(X) & $\log\epsilon$\,(X) 
                                      & $\mbox{[X/Fe]}$\xx & $\sigma$ & $N$ &
                                      & $\log\epsilon$\,(X) 
									  & $\mbox{[X/Fe]}$\yy & $\sigma$ & $N$ }
\startdata 
\ion{C}{1}  & 8.43 &     7.10 &  $+$1.08\zz &    0.20 &  2 && \nodata &  \nodata & \nodata & \nodata \\
O (OH)      & 8.69 &     6.60 &  $+$0.32    &    0.30 &  1 && \nodata &  \nodata & \nodata & \nodata \\
\ion{Ti}{1} & 4.95 &     2.98 &  $+$0.44    &    0.04 &  2 && \nodata &  \nodata & \nodata & \nodata \\
\ion{Ti}{2} & 4.95 &     2.96 &  $+$0.42    &    0.19 & 11 && \nodata &  \nodata & \nodata & \nodata \\
\ion{Cr}{1} & 5.64 &     3.35 &  $+$0.12    &    0.31 &  4 && \nodata &  \nodata & \nodata & \nodata \\
\ion{Cr}{2} & 5.64 &     3.17 &  $-$0.06    &    0.07 &  3 && \nodata &  \nodata & \nodata & \nodata \\
\ion{Mn}{2} & 5.43 &     2.80 &  $-$0.22    &    0.15 &  2 &&    3.50 &  $-$0.42 &    0.20 &       2 \\
\ion{Ni}{1} & 6.22 &     3.78 &  $-$0.03    &    0.20 & 14 && \nodata &  \nodata & \nodata & \nodata \\
\enddata
\tablenotetext{a}{\metal=$-$2.41}
\tablenotetext{b}{\metal=$-$1.51}
\tablenotetext{c}{\cfe$>+$1.31 using corrections of \citet{placco2014c}.}
\end{deluxetable}

\begin{deluxetable}{lrrrr}
\tabletypesize{\scriptsize}
\tablewidth{0pc}
\tablecaption{Systematic Abundance Uncertainties\label{sys}}
\tablehead{
\colhead{Elem}&
\colhead{$\Delta$\teff}&
\colhead{$\Delta$\logg}&
\colhead{$\Delta\xi$}&
\colhead{$\sigma_{\rm tot}$}\\
\colhead{}&
\colhead{$-$150\,K}&
\colhead{$-$0.5 dex}&
\colhead{$-$0.3 km~s$^{-1}$}&
\colhead{(dex)}}
\startdata
\multicolumn{5}{c}{HD~196944} \\
\hline
\ion{Fe}{1} & $-$0.18 & $-$0.01 & $+$0.03 & 0.18 \\
\ion{Fe}{2} & $-$0.06 & $-$0.19 & $+$0.03 & 0.20 \\
\ion{Na}{1} & $-$0.17 & $+$0.14 & $+$0.14 & 0.26 \\
\ion{Mg}{1} & $-$0.16 & $+$0.06 & $+$0.03 & 0.17 \\
\ion{Ca}{1} & $-$0.12 & $-$0.01 & $+$0.00 & 0.12 \\
\ion{Sc}{2} & $-$0.14 & $-$0.13 & $-$0.06 & 0.20 \\
\ion{Ti}{1} & $-$0.24 & $-$0.05 & $+$0.06 & 0.25 \\
\enddata
\end{deluxetable}

\end{document}